\documentclass{aa}  

\usepackage{graphicx}
\usepackage{color}
\usepackage{txfonts}
\newcommand{\el}[1]{\textcolor{black}{#1}}
\newcommand{\red}[1]{\textcolor{black}{#1}}
\renewcommand\makeLineNumber{}

\begin{document} 
\nolinenumbers

\title{On the formation of multiple dust-trapping rings in the inner Solar system}
   \subtitle{}

   \author{ Lega E.\inst{1},  Morbidelli A.\inst{1,2},  Masset, F. \inst{3} and B\'ethune,W.\inst{4} }

\institute{ Université Côte d'Azur, Observatoire de la Côte d'Azur, CNRS, Laboratoire Lagrange, France\label{inst1}
   \and
    Collège de France, 11 Pl. Berthelot, 75005 Paris, France \label{inst2}
      \and
   Instituto de Ciencias Físicas, Universidad Nacional Autónoma de México, Av. Universidad s/n, 62210 Cuernavaca, Mor., Mexico \label{inst3}
   \and
   DPHY, ONERA, Université Paris Saclay, F-91123 Palaiseau, France \label{inst4}
   }

   \date{Received S; accepted }

\abstract
{ Isotopic properties of meteorites provide evidence that multiple dust trap or pressure bumps had to form and persist in the inner Solar System on a timescale of millions of years. The formation of a pressure bump at the outer edge of the gap opened by Jupiter is effective in  blocking particles drifting from the outer to the inner disk. But this is not enough to preserve dust in the inner disk.  However, in low viscosity disks and under specific condition on the gas cooling time, it has been shown that massive planets can also open secondary gaps, separated by density bumps, inward of the main gap. The majority of studies of the process of secondary gap formation have been done in two dimensional equatorial simulations with prescribed disk cooling or approximating the cooling from the disk photosphere.  Recent results have shown that an appropriate computation of the disk cooling by including the treatment of radiation transport is key to determine the formation of secondary gaps.}
{  Our aim is to extend previous studies to three dimensional disks by also including
radiative effects. Moreover, we also consider non ideal MHD effects, in disks with prescribed cooling time, to explore the feedback of the magnetic field on secondary gap formation.}
{ We perform three dimensional hydrodynamical numerical simulations with self consistent treatment of radiative effects making use of the Flux Limited Diffusion (FLD) approximation. We then extend our study to a similar disk  including  the magnetic field and the non ideal Ohmic and Ambipolar effects.}
{We show that, in the hydrodynamical model, a disk with low bulk viscosity ($\alpha_{\nu}=10^{-4}$) and consistent treatment of radiative effects, a Jupiter massive planet opens multiple gaps. We also show that multiple gaps and rings are  formed by planetary masses close to the pebble isolation mass. In the presence of non ideal MHD effects multiple gaps and rings are also formed by a Jupiter massive planet.}
{A  Jupiter solid core  in low viscosity disks blocks particles drifting towards and within the inner disk. The formation of multiple gaps and rings inside the planetary orbit at this stage is crucial to preserve dust reservoirs. Such reservoirs are pushed towards the inner part of the disk during Jupiter runaway growth {\el {and they are shown to be persistent  after Jupiter's growth}}. Multiple dust reservoirs could therefore be present in the inner Solar System  since the formation of Jupiter's solid core if the disk had low-viscosity. }

\titlerunning{Multiple dust-trapping rings}

\authorrunning{Lega E.} 

\maketitle

\section{Introduction}

Meteorites divide in two broad categories: chondrites and achondrites, the latter including iron meteorites. The parent bodies of achondrites formed in the first My of Solar system life. Thus, they contained a large abundance of short-lived radioactive elements, whose decay liberated enough energy to melt the body \citep{Neumann2012}. Under the effect of gravity, these bodies then differentiated into an iron core, a silicate mantle and a crust, akin of asteroid 4 Vesta. The parent bodies of chondrites instead formed sufficiently later, so that most short-lived radioactive atoms had already decayed. Consequently, the energy liberated in their interior was much less; temperature increased significantly, up to several hundred degrees, but not enough to cause melting. Thus, these bodies preserved their original petrological structure, made of chondrules, matrix and refractory inclusions. From the measurements of the ages of formation of the individual chondrules, the chondrite parent bodies formed approximately 2 to 4 My after the beginning of the Solar system \citep{Fukuda2022, Piralla2023}. 
The formation of planetesimals so late in the history of the protoplanetary disk is problematic, because dust particles (i.e. the individual chondrules, or the matrix grain aggregates) tend to drift radially towards to Sun on a timescale much shorter than a million years. One could think that the dust lost by radial drift at a given location was substituted by dust drifting from larger radial distances in the disk. But this simple view is negated by another crucial observation in meteorite science. The analysis of nucleosynthetic isotopic anomalies inherited from the interstellar medium reveals that both chondrites and achondrites separate in two distinct groups, dubbed CC (because the measured anomalies are like those of carbonaceous chondrites) and NC (because the anomalies are like those of non-carbonaceous chondrites) \citep{Warren2011,Kruijer2017}. 
The existence of an isotopic dichotomy like that described above implies that the objects of the NC and CC groups formed at different locations of the disk, incorporating distinct materials. Because carbonaceous chondrites show evidence of prominent water alteration, whereas the isotopic properties of NC meteorites are similar to those of Earth and Mars, the common interpretation is that the NC bodies (both achondrites and chondrites) formed in the inner Solar system and the CC bodies (again both achondrites and chondrites) formed in the outer Solar system where water could be accreted in solid form \citep{Kruijer2017,Nanne2019}. An implication of these considerations is that the material that formed the NC chondrites in the inner Solar system at a late time could not have drifted from the outer disk, otherwise it would have had CC isotopic properties. The formation of Jupiter, which opened a gap in the disk \citep{Kruijer2017} or the formation of a pressure bump near the snowline \citep{Brasser2020} must have blocked the drift of particles from the outer disk to the inner disk. Again, meteorites provide evidence that only particles smaller than 200 microns could pass through such dynamical barriers, with negligible effect on the overall mass balance \citep{Haugbolle2019}. \par
Thus, in absence of a refilling mechanism, the NC dust in the inner part of the protoplanetary disk had to survive for millions of years without drifting into the Sun. This requires the existence of a pressure bump in the disk where dust could be trapped. 
A pressure bump may be generated near 1 au if the disk evolves primarily under the effects of magnetized winds, and if the mass removal rate in these winds increases with decreasing heliocentric distance \citep{Suzuki2010,Ogihara2018}. The existence of a unique pressure bump, however, is not completely satisfactory {\el {\citep{2020SSRv..216...55K,2020E&PSL.55116585S}}}. Four kinds of NC chondrites exist, with somewhat different isotopic anomalies: the prominent Ordinary and Enstatite chondrite groups, and the Rumuruti and Kakangari groups, less represented in the meteorite collections.  This, together with the fact that Enstatite chondrites have essentially identical isotopic properties of a class of NC achondrites called Aubrites, implies that dust did not completely mix up, even on a timescale of millions of years.  This strongly suggests that not one, but several pressure bumps had to form and persist in the inner part of the protosolar disk.  \par
Multiple rings of dust are observed in protoplanetary disks \citep{Andrews2018}, suggestive of a multitude of pressure bumps. It is debated whether the formation of these pressure bumps is due to the formation of massive planets opening gaps in the disk \citep{Bae2017} or to spontaneous non-ideal MHD effects  \citep{Bethune2017, Riols2018, Cui2021}. We remark that, spontaneous generation of rings has been obtained (by  \cite{Bethune2017, Riols2018, Cui2021}) in cases where ampipolar diffusion is the dominant non ideal effect. 
This result suggests that such rings should not form in the inner disk because the MHD dynamics there is thought to be dominated by Hall and Ohmic dissipation, rather than ambipolar diffusion \citep{Armitaage2011}. \par
The rings are observed only in the outer part of protoplanetary disks and, due to resolution effects, it is not possible to assess observationally whether they are common also in the inner part.
 Also, none of the terrestrial planets was massive enough to open a gap in the disk, even if already present with its current mass, because their Hill sphere is much smaller than the vertical scale height of the disk.  
Therefore, Jupiter remains the main suspect for the potential formation of rings in the inner disk. Jupiter obviously forms a prominent pressure bump outside of the main gap that opens along its orbit, but this is of no help for the preservation of dust in the inner  system. It has been shown, however, that massive planets can also open secondary gaps, separated by density bumps, also inward  of the main gap  \citep{Bae2016,Bae2017,Miranda2020}.

\par Most studies of the process of secondary gap formation have been conducted in 2D equatorial simulations \citep{Bae2017,2018ApJ...869L..47Z,2019ApJ...875...37M,Miranda2020,Ziampras2023} and show that three necessary conditions have to be met: (i) the disk must have a very small effective turbulent viscosity; (ii) the disk cooling time normalized by the local orbital frequency, denoted $\beta_{c}$, has to be either small ($\beta_{c}<0.01$)  or large ($\beta_{c} >10$) \citep{Miranda2020,Ziampras2023}; (iii) the planet must be massive enough (the mass depending on the disk’s viscosity, $\beta_{c}$ and on disk’s scale-height.
\par
Including a prescription of radiative effects gives the appropriate cooling time associated to disk properties  and is a
necessary step to capture the dynamics of
the formation of secondary gaps in specific disks.
Taking into  accounts the cooling from the disk surface \citep{Ziampras2020} and adding radiation transport along the disk midplane
 \citep{Miranda2020,Ziampras2023} have improved the treatment of disk thermodynamics in two dimensional simulations. However, a self-consistent treatment of  radiative
 effects in three dimensional simulations of secondary gap opening has not yet been considered.
To our knowledge, the only 3D simulations with an appropriate $\beta_{c}$ from disk surface irradiation and a Jupiter-mass planet in a massive, inviscid protoplanetary disk was presented in \cite{Bae2016}, which showed several waves in the disk’s surface density distribution inward of Jupiter’s orbit. The authors, however, did not perform an analysis of the pressure profile in the disk to verify whether dust-trapping pressure bumps formed. 
\par
Given the relevance of this problem to understand the formation of different classes of NC chondrites, we present an exploration of the capability of Jupiter to form multiple dust-trapping rings in the inner Solar system using a self-consistent treatment of radiative effects in a disk with low bulk viscosity.
\par
Moreover, since low viscosity disks  transport gas towards the star at a very low rate with respect to  typical accretion rates  observed in young stars \citep{1998ApJ...495..385H,2016A&A...591L...3M} we extend our model by including non-ideal magneto hydro dynamics .  In this case, accretion flow onto the star is obtained by angular momentum removal by disk winds \citep{Bethune2017,Lesur2021} and our aim is to explore the feedback of the magnetic field on secondary gap formation. \par
The paper is organized as follows:
in Section 2 we present our simulation tools and methods; in Section 3 we reproduce the work of \cite{Bae2016} and compute the pressure profile on the midplane. In Section 4 we present a different simulation, where we adopt a density  comparable to the minimum mass solar nebula at the planet location, instead of the massive disk used by \cite{Bae2016}, and we compute $\beta_{c}$ self-consistently using flux-limited energy diffusion. 
In Section 5 we extend our model including non-ideal magneto hydrodynamics.
Finally, in Section 6 we progressively reduce the mass of the planet to see at which mass a planetary core might have started to form dust-trapping rings in the inner Solar system.  

\section{Models}
\subsection{Physical model}
\label{Sec:model}
The protoplanetary disk (PPD hereafter) is treated as a non-self-gravitating gas whose motion is described by the Navier-Stokes equations. We use two  grid-based  codes: the \textit{fargOCA} code \citep{2014MNRAS.440..683L} which integrates a full energy equation taking into account viscous heating and radiative cooling and the \texttt{FARGO3D} code \citep{Benitezetal2016} which includes non-ideal MHD effects. The two codes are very similar in the sense that they use the same hydrodynamical solvers based on the operator splitting procedure  \citep{StoneandNorman1992} and they  both  benefit  from the  \texttt{FARGO} algorithm \citep{Masset2000} for a computation of the time-step  suited to  gas rotating with  quasi-Keplerian motion.
The reason for the use  of the two codes solely depends on the different physical modules that they implement.\par
In both cases we use spherical coordinates $(r,\theta,\varphi)$ 
where $r$ is the radial distance from the star (which is at the origin of the coordinate system), $\theta$ the 
polar angle measured from the $z$-axis (the colatitude), and $\varphi$ is the azimuthal coordinate measured from the $x$-axis. The midplane of the disk is located at the equator
$\theta = \frac {\pi} {2}$. 
We work in a coordinate system which rotates with angular velocity of a planet of mass $m_p$:
$$\Omega_k = \sqrt {\frac{G(M_{\star}+m_p)}{{r_p}^3}} $$
where $M_{\star}$ is the mass of the central star, $G$ is the gravitational constant,
and $r_p$ is the semi-major axis of  a planet  assumed to be on a circular orbit.
We consider a planet orbiting a Solar mass star and therefore in the following we will indicate the
mass of the star as $M_{\sun}$.

\par
The dynamics of the gas is provided by the integration of the  Navier-Stokes equations composed by a continuity equation and a set of 3 equations for the momenta \citep{2014MNRAS.440..683L,Benitezetal2016}.
Additionally, we consider the gas internal energy density $e=\rho c_v T$, where $\rho$
and $T$ are the gas volume density and the gas temperature  and
$c_v$ is the specific heat at constant volume. The equation of state is that of an ideal gas of pressure 
 $P = (\gamma-1)e$  with  adiabatic index $\gamma=1.4$.
In the following we use either of two prescriptions for the integration of the energy equation:
\begin{enumerate}
\item
	Changes in the internal energy {\el {density $e$}} are due to adiabatic (de)compression, viscous heating  \citep{MihalasMihalas84} and a relaxation  to the initial temperature $T_0$ on a  cooling timescale ${\tau_{\rm c}}= \beta_c \Omega_k^{-1}$:
\begin{equation}
\label{eq:EdotRelax}
{\partial_t   e } + \nabla \cdot(e\vec {\rm v})  =  -P \nabla \cdot \vec {\rm v} +Q^+ -\rho c_v\frac{\left(T-T_{\rm 0}\right)}{\tau_{\rm c}}, 
\end{equation}
where $\vec {\rm v}$ is the gas velocity, and the terms $-P \nabla \cdot \vec {\rm v}$ and $Q^{+} \equiv \bar{ \bar \tau}\nabla v$ are respectively the compressional and viscous heating
{\el { ($\bar{\bar \tau}$ is  the viscous stress tensor defined in \citet{MihalasMihalas84})}}. 
The dimensionless cooling time $\beta_c$ is either a constant parameter or a function depending on disk properties as will be described below. 
 \item
  We  consider  the evolution of the internal energy  and of the thermal radiation energy $E_r$  (code {\it fargOCA}) in the 
  flux-limited diffusion approximation \citep[FLD,][]{LevermorePomraning81} using the so-called two temperature approach \citep{Commercon2011}:
\begin{equation}
\label{eq:Edot2Temp}
\left\lbrace \begin{array}{lll}
  {\partial_t E_{\rm r}} - \nabla \cdot  \vec F & = &
    \rho \kappa_p c(a_rT^4 - E_{\rm r}) \\
  {\partial_t e} +  \nabla \cdot(e\vec {\rm v}) &= 
&
  -P \nabla \cdot \vec v -\rho \kappa_p c(a_rT^4 - E_{\rm r})  + Q^+
\end{array} \right.
\end{equation}
where  $\vec F =   \frac{c\lambda}{\rho \kappa_r} \nabla E_{\rm r}$ is the radiation flux vector and $\lambda$ the flux limiter \citep{KBK09}.
We indicate with $\kappa_p$ and  $\kappa_r$   respectively the Planck  and the Rosseland mean opacity, with $a_r$  the radiation constant and with $c$  the speed of light.
In this paper we consider $\kappa = \kappa_p \equiv \kappa_r$ \citep[see][]{Bitschetal2013} and use the opacity law of \cite{1994ApJ...427..987B} \red{where a dust to gas ratio of 0.01 is assumed.}
 \par 
 
 We do not include the irradiation heating from the central star in order to  reduce the computational cost with no significant impact on the gas dynamics in the planet vicinity \citep[see][]{Legaetal2015}.
  However, we consider in this work low viscosity disk
   having the stellar irradiation as the main source of heating.
Therefore, in our three dimensional simulations, we  mimic the thermal equilibrium structure obtained in disks heated by the  star, by suitably fixing the disk surface temperature (see Section \ref{Sec:rad} and \cite{2024A&A...690A.183L}). \par
 When considering the energy evolving according to Eq.\ref{eq:Edot2Temp}  the disk cooling time $\beta_c$ 
 fully depends on disk properties and can be estimated a posteriori as explained below. In the following we will call "fully radiative" the simulations  with the above  prescription for energy computation.   
 \end{enumerate}

 \subsection{Cooling time approximation}
 When computing an energy equation with temperature relaxing towards the initial values (Eq.\ref{eq:EdotRelax}) it is customary to consider the cooling time $\beta_c$   as a constant parameter  (see for example \citep{2020ApJ...892...65M,Ziampras2023}). However, when studying specific disks, or in the aim of comparison to observed disk structures, it is important to compute the cooling time from disk properties \citep{2018ApJ...869L..47Z,Miranda2020,Ziampras2023}. 
 \subsubsection{Cooling through disk surfaces}
 The cooling process occurs through blackbody emission of dust grains. 
 In vertically integrated disks the radiative  cooling \el{ power per unit surface} is given by \citet{2004ApJ...606..520M}:
 $$Q_{cool} = - \frac{\sigma T^4}{\tau_{eff}}$$
 where $\sigma= a_rc/4$ and $\tau_{eff}$ is the effective optical depth
 \citep{1990ApJ...351..632H}:
 $$\tau_{eff}=\frac {3}{8}\tau+\frac {\sqrt 3}{4}+\frac {1}{4\tau} $$ to take into account optically thick and thin limits, and $\tau=1/2 \kappa \Sigma $ is the optical depth at disk midplane.
 
 The disks cools through surfaces and, following \citet{Ziampras2020} the surface cooling time
 can be written as:
\begin{equation}
\label{beta_surf}
    \beta_c^{surf} =  {\Omega_K} {\frac{\Sigma c_v T}{|2 Q_{cool|} }}\equiv  {\Omega_K}{\frac{\Sigma c_v\tau_{eff}}{2\sigma T^3 }}
\end{equation}
 \el {The formula has been extended to a  three dimensional disk  by   \cite{2016ApJ...817..102L} by considering the radiative timescale  $t_r=\frac {e}{\nabla \cdot \vec F}$ (with $e$ and $\vec F$ defined in Eq.\ref{eq:Edot2Temp}) and integrating over a spherical volume of radius $H$} \footnote{The integral form is preferred to avoid the computation of the divergence}:
\begin{equation}
      \label{BaeTauCooling}
       \beta_c^{H} = {\Omega_K}\frac{  \rho c_v H \tau_{eff}}{3 \sigma T^3}.
   \end{equation}
  In this case the  optical depth  $\tau_{eff}$ is computed for grid cells at height  $z$ above (below) the midplane  as:  $\tau(z)  = \int _z ^{z_{max}} \rho(z') \kappa(z') dz'$ 
 $\left( \tau(z)  = \int _{z_{min}} ^{z} \rho(z') \kappa(z') dz'\right)$.

 \begin{table*}
\begin{minipage}{160mm}
\caption{Disk models parameters.}
\label{table:tab1} 
\begin{tabular}{|lllllllll|}
\hline\noalign{\smallskip}
Name  &  $\Sigma_0/ M_{\sun}r_0^{-2}$  & $\alpha_\Sigma$ & $(H/r)_{r_p}$ & $r_{min}/r_0$ &  $r_{max}/r_0$ & $\theta_0$ & $\alpha_{\nu}$ & $(N_r,N_\varphi,N_\theta)$ \\
\noalign{\smallskip}\hline\noalign{\smallskip}
\hline\noalign{\smallskip}
HD$_{Bae}$ & $2.3 \, 10^{-3}$ & 1.5 & 0.05 & 0.2 & 1.5 & 1.36 & 0 & $(726,754,72)$\\
HD$_{rad}$J, HD$_{rad}$S, HD$_{rad}$30  &   $6.67 \, 10^{-4}$ & 1.07 & 0.03 & 0.3 & 3.5 & 1.48 & $10^{-4}$ & $(528,554,52)$ \\
HD$_{iso}$, HD$_{\beta}$, HD$_{adia}$ &   $6.67 \, 10^{-4}$ & 1.07 & 0.03 & 0.2 & 3.5 & 1.48 & $10^{-4}$ &  $(658,650,62)$  \\
MHD & $1.3 \, 10^{-4}$ & 1 & 0.05 & 0.3 & 10.0 & 0.56  & 0 & $(384,800,64)$ \\\\
\hline
\end{tabular}
\tablefoot{In the MHD case, we run 6 simulations by varying the intensity of the magnetic field and the strength of the Ambipolar diffusion. The MHD simulations share the same disk parameters as defined in the table, therefore we keep a unique name and explicitly provide the values of  $\Lambda_A$ and of the strength of the magnetic field in Section 5.}
\end{minipage}
\end{table*}

 \begin{table}
\begin{minipage}{85mm}
\caption{Boundary conditions}
\label{table:tab2} 
\begin{tabular}{|lll|}
\hline\noalign{\smallskip}
Name   & Radial & Vertical   \\
\noalign{\smallskip}\hline\noalign{\smallskip}
\hline\noalign{\smallskip}
HD$_{Bae}$   & Damping   & Open \\
HD$_{rad}$   & Damping & Reflecting$^{1}$ \\
HD$_{iso}$ , HD$_{\beta}$, HD$_{adia}$  & Damping & Reflecting  \\
MHD         & see Appendix\ref{App:MHDBound}  &   see Appendix \ref{App:MHDBound}  \\\\
\hline
\end{tabular}
\tablefoot{The boundaries in the table refer only to gas fields. 
\el {$^{1}$ In the fully radiative case the temperature at the disk surface is provided by the equilibrium between stellar heating and radiative cooling
(see Section \ref{Sec:rad}). This equilibrium temperature is used for the computation of both the internal and  the radiative energy densities.} In the MHD simulations additional boundary conditions for the magnetic field and the emf are required. We report the details in Appendix \ref{App:MHDBound}. All the simulations presented in this paper run on one single hemisphere. Therefore, the vertical boundaries refer to disk's surface. On midplane we use mirror boundary conditions.}
\end{minipage}
\end{table}

 \subsubsection{In-plane cooling}
  Dust grains emission actually occurs in all directions, i.e the radiative flux has also a radial component in addition to the previously described surface cooling. A detailed study on the role of the radial radiative flux or in-plane cooling on the evaluation of the cooling timescale has been done by \cite{Miranda2020} and \cite{Ziampras2023}  on two dimensional vertically integrated disk models. In those papers, an effective cooling time scale is obtained by combining the surface cooling of Eq.\ref{beta_surf}
  and the in-plane cooling obtained  considering  heat diffusion  through the mid-plane by  \cite{Miranda2020}. The results obtained underline the importance of considering the in-plane cooling component and suggest that a study of  three dimensional disks with heat diffusion from the FLD approximation is required to evaluate thermal processes and correctly estimate  the cooling time $\beta_c$.
  
  \subsection{Cooling time in three dimensional disks with FLD approximation}
  When dealing with Eq.\ref{eq:Edot2Temp}  heat diffusion is fully taken into account by computing the radiation flux on every grid-cell at any time $t$.\par
  The cooling time can therefore be derived a posteriori  from the timescale over which radiation is diffused over a typical length scale $l_r$:
\begin{equation}
    \label{Eq:tau}
    \beta_c =\Omega_K \frac {l_{r}^2} {D_r}
\end{equation}
where $D_r$ is the radiative diffusion coefficient of a grid cell at time $t$. In the optically thick limit the diffusion coefficient associated with the radiation flux vector in Eq.\ref{eq:Edot2Temp} is: 
\begin{equation}
\label{Eq:diffOt}
    D_r = \frac {16\sigma T^3} {3c_v\rho^2\kappa}
\end{equation}
where we have considered $E_{\rm r} = a_r T^4 $ and $\lambda =1/3$. 
Although Eq.\ref{Eq:tau} overestimates the cooling time in the optically thin regions, we keep this expression   since  typical disks are optically thick  in the planet's forming region near the midplane. \par
By considering the disk scale height $H$ as the characteristic  diffusion  length scale , we obtain ( see also \cite{Miranda2020,Ziampras2023}):
\begin{equation}
    \label{Eq:tauOt}
    \beta_c =\Omega_K \frac {3c_v\rho^2\kappa H^2} {16 \sigma T^3}
\end{equation}
We will compute this quantity  to compare  results obtained with the energy evolving with FLD approximation to those obtained  with eq.\ref{eq:EdotRelax}.
 We remark that  the cooling time obtained from diffusion
 Eq. \ref{Eq:tauOt} in optically thick disk is similar to  the cooling from disks photosphere as given in Eq. \ref{beta_surf} and Eq. \ref{BaeTauCooling}.
 
 \subsection{Non ideal magnetic effects}
 In order to include MHD effects (with the code \texttt{FARGO3D})  we also consider the induction equation for the evolution of the magnetic field $\vec B$:
 \begin{equation}
\label{MHDequa}
{\partial_t  \vec B }  =  \nabla \times (\vec v \times \vec B -\eta _O \bf J + \eta _A \bf J \times \vec e_b \times \vec e_b)
\end{equation}
With respect to the pure hydrodynamical case the equation for the momenta have an additional source term: $\vec J \times \vec B$ \citep{Benitezetal2016} where $\vec J \equiv \nabla \times \vec B$
is the electric current.  \par
The terms  $\eta_O$ and $\eta_A$ are the Ohmic and ambipolar diffusivities and $\vec e_b$ is the unit vector parallel to the magnetic field line. We do not consider  the Hall effect.
\el{ There are multiple reasons for this choice: i)  Ohmic resistivity is considered dominant in the region of interest near the midplane where the density is large, and recently it has been supposed that it is dominant even at low density values \citep{ 2024arXiv240506026H}; ii)  to our knowledge  spiral arms' propagation in MHD disks has not yet been investigated so that we choose a relatively simple setting, iii) finally, from the computational point of view  3 dimensional MHD simulations with embedded planets are extremely expensive and Hall effect can contribute to  further decreasing the time step which would make prohibitive the  integration of the system for about 100  planetary orbits.}\par
Following \cite{Lesur2021} and \cite{Wafflard2023} we prescribe the Ohmic and ambipolar profiles instead of considering a model for ionization that would also make the three-dimensional computations extremely  expensive. 
We consider the dimensionless Ohmic Reynolds number $\mathcal R_O$ and Elsasser number $\Lambda _A$ defined \el {at disk midplane} by:

$$\Lambda _A(0) \equiv  \frac{v_A^2}{\eta_A \Omega_k}  \,\,\, \mathcal R_O(0) \equiv \frac{\Omega_K H^2}{\eta_O }$$

where $v_A = B/\sqrt{\rho \mu_0}$ is the Alfv\'en velocity.
According to \cite{Lesur2021} and \cite{Wafflard2023}, we consider that the gas is fully ionized in the  corona. This is set through:
$$\Lambda_A(z)= \max\left\{\Lambda_A(0)\exp \left[ \frac{z^4} {(\xi H)^4}\right],\frac{1}{10}\frac{v_A^2}{c_s^2}\right\}$$
$$ \mathcal R_O(z) = \mathcal R_O(0) \exp \left[ \frac{z^4} {(\xi H)^4}\right]\left(\frac{\rho(0)}{\rho(z)}\right)$$
where $\xi$ quantifies the thickness of the non-ideal layer in units of the pressure scale height $H$ ($\xi =3$ in our simulations).
\par
We notice that the models that compute the Ohmic and ambipolar coefficients taking into account the ionization of gas and chemical models  \citep{Gressel2015,Bethune2017,Bai2017,Lesur2021,Lesur2021b} do not provide a clear consensus on the values of the strength of non ideal effects. This is mainly due to a strong uncertainty about the ionization rate from cosmic rays (the main source of ionization below two scale height) and on the size and abundance of grains in disks that tend to reduce the ionisation fraction.\par
In the simulations presented in Section \ref{Sec:MHD} we consider $\mathcal R_O = 1 $ and different values of  $\Lambda _A \in [0.01:10] $.
This choice is motivated by the fact that values of $R_{O}\sim {\it O}(1)$ and  strong ($\Lambda_A \sim 0.01$)  to weak  ambipolar diffusion ($\Lambda_A \sim 10$) are plausible \citep{Gressel2015,Bethune2017,Bai2017} in the region of the protoplanetary disk considered in this paper.

 \subsection{Disk models}
 
 The code units are such that $G=M_*\equiv M_\odot=1$, and the unit of distance $r_0$ is arbitrary.
The unit of time is therefore $(r_0/{\rm au})^{3/2}/(2\pi)\,{\rm yr}$. 
For the simulations we adopt as the unit of length $r_0=5.2$~au except for the MHD case for which $r_0=1$~au. Planets are considered, in all cases, at position $r_p=5.2$ au.
We call $T_0$ the orbital period at $r=r_0$.
The simulations' domain extend radially from $r_{min}$ to $r_{max}$ and vertically 
from the colatitude  $\theta_0$ to the midplane at $\theta = \pi/2$. 
We consider  disks 
with radial profile of surface density $\Sigma (r) = \Sigma_0 (r/r_0)^{-\alpha_\Sigma}$. 
The values of these parameters as well as simulation names are reported in table \ref{table:tab1}. 
\el {The fully radiative simulations are named HD$_{rad}$ followed by $J$, $S$ and $30$, for
planets of respectively the mass of Jupiter, of Saturn  and of a  super Earth of $30$ Earth masses.}
We consider non zero $\alpha$ viscosity described by the parameter $\alpha_{\nu}$ for the whole set of hydro-dynamical simulations except HD$_{Bae}$. In the MHD simulations angular momentum is transported by the magnetic field without the need to prescribe a viscosity (e.g., of turbulent
origin).

Other parameters like the boundary conditions and the numerical resolution do also depend on the simulation and are specified in table \ref{table:tab2}. In the azimuthal direction we use periodic boundary conditions. The label "Damping" refers to the implementation of a wave damping region according to \cite{deValBorroetal2006}.  The radial grid spacing is logarithmic, the azimuthal one is constant, while the vertical spacing is  constant except for the MHD  simulations for which we use a nonuniform grid. The reason is that magnetic effects like winds develop on a large vertical domain and a nonuniform grid  meet the requirements of having a  moderate total number of grid cells together with a suited resolution  ($\sim 8$ grid cells) over a disk scale height above the midplane. \el {Precisely, our grid has constant resolution from the midplane up to one scale height, and then it varies quadratically with a maximum resolution ratio of four}. Moreover, in the case of MHD simulations  additional boundary conditions are required for the magnetic field and for the electromotive field (EMF). We report the details in Appendix \ref{App:MHDBound}.

 \subsection{Disks with embedded planets: numerical procedure}
In all the simulations we consider a planet on a fixed circular orbit located  on the midplane at $r_p = 5.2$ au (i.e $r_p=r_0$ for all the simulations except the MHD ones for which $r_p=5.2 r_0$ with $r_0=1$ au) and azimuth $\varphi = 0$, or equivalently at $(x_p,y_p,z_p)=(5.2,0,0)$ au.
We embed in each disk model a planet of $20$ Earth masses at $t=0$ and smoothly increase, over a time interval of 10 or 20 orbits, the mass until the final mass is reached. The fully radiative and MHD cases   require   preliminary two dimensional
($r,\theta)$ simulations to reach an equilibrium configuration before planet insertion. This phase will be described in the corresponding sections.
\par
 The gravitational potential of the planet acting on
the disk ($\Phi_p$) is modelled as in \citet{KBK09} in the {\it{fargOCA}} code: the full gravitational potential is computed  for disk elements having distance $d$ from the planet larger than a fraction $\epsilon$ of the Hill radius,
while  it is  smoothed  for disk elements with $d<r_{sm} \equiv \epsilon R_H$ according to:
\begin{equation}
    \label{KleyPotential}
\Phi _p = \left\lbrace \begin{array}{ll}
-\frac {m_pG} { d} &  d > r_{\rm sm} \\
-\frac {m_pG} { d}f(\frac{d}{ r_{\rm sm}}) & d\leq r_{\rm sm}   
\end{array} \right .
\end{equation}
with $f(\frac {d}{r_{\rm sm}}) = \left [ \left( \frac{d} {r_{\rm sm}}\right)^4-2\left( \frac {d}{ r_{\rm sm}}\right)^3+2\frac{d}{ r_{\rm sm}} \right]$.\par
In \texttt{FARGO3D} the potential is smoothed according to:
\begin{equation}
\Phi _p = -\frac {m_pG}{\sqrt{(d^2+r_{sm}^2)}}
\end{equation}
We recall that the Hill radius is defined as:
$R_H = r_p(m_p/3M_{\sun})^{1/3}$. 
The values of $\epsilon$ are respectively
$(0.13,0.2,0.1)$ in simulation sets (HD$_{Bae}$,HD,MHD).

\begin{figure}
\centering
\includegraphics[width=\hsize]
   {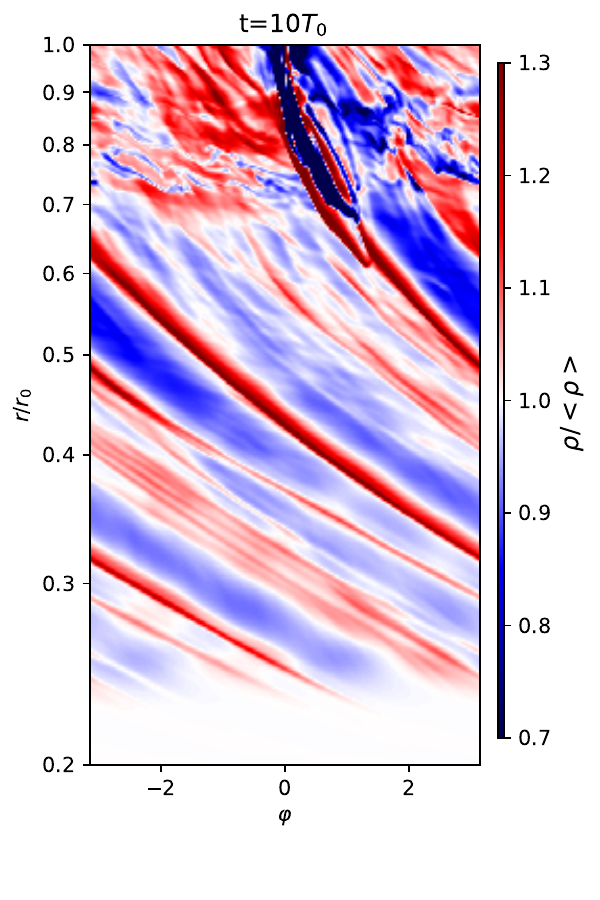}
      \caption{Simulation HD$_{Bae}$. Two dimensional distribution in the $(r,\varphi)$ plane of the midplane density $\rho$ normalized over the azimuthally averaged value $<\rho>$ at $t=10$ orbits when the planet has just reached a Jovian mass. In order to compare to Fig.6 of \citet{Bae2016} the y-axis is plotted using a logarithmic scale.}
         \label{Fig:BaeFig6} 
\end{figure}

\section{Hydrodynamical case with prescribed  cooling}
\label{Sec:bae}

The only three dimensional simulation, to our knowledge, which shows the formation of multiple rings and gaps inside the orbit of a Jupiter mass planet is the one in \citet{Bae2016}. This paper was mainly devoted to the study of spiral waves induced by a massive planet, however it offers  an important starting point for our study.
The aim of this section is to produce a simulation similar to the one presented in Section 4.2 of \citet{Bae2016} and to check whether the density maxima that form inside the Jupiter's orbit correspond to pressure bumps with the ability of stopping dust from drifting inward. \par
Although we do not consider dust dynamics in our simulations, we notice that dust velocity would be affected by the gas drag; precisely, introducing a radial dust velocity $u_r$ we have \citep{1986Icar...67..375N,2002ApJ...581.1344T}:
\begin{equation}
\label{dust_vr}
    u_r = \frac{v_r}{1+S_t^2} -2S_t\eta v_K
\end{equation}
where $S_t$ is the Stokes number, i.e. the value of the friction timescale in units of the orbital frequency,
and $\eta$ is the fraction of the Keplerian velocity $v_K$ corresponding to the headwind experienced by dust particles:
\begin{equation}
\label{eta}
\eta = -\frac{1}{2}\left(\frac{H}{r}\right)^2 \frac {\partial \log P}{\partial \log r}
\end{equation}
Neglecting the gas radial velocity $v_r$, which is expected to be small, a negative derivative of the pressure with respect to $r$ corresponds to  $\eta >0$ so that  the dust moves toward the star,
while for a positive pressure gradient  ($\eta <0$) the situation is reversed. A dust trap is located at the pressure maximum where  $\eta =0$.\par
In locally isothermal disks  $P=c_s^2\rho$,
while for more generic Equations of State (EoS) $P=(\gamma-1)e$.
The disk that  we consider in simulation HD$_{Bae}$ is initialised with a locally isothermal pressure, which gives a   value of $\eta$
in the unperturbed disk  of $~0.003$ (considering the  parameters in table \ref{table:tab1}).
\par
 In our simulation HD$_{Bae}$ we consider a non viscous disk and  Eq.\ref{eq:EdotRelax} for the time evolution of the internal energy density as in \cite{Bae2016}.
 However, we do not implement the computation of $\beta_c^{surf}$ from Eq.\ref{BaeTauCooling}  and we consider a constant prescribed $\beta_c$ 
corresponding to   $\beta_c \sim  100 \sim \beta_c^{H}$
(see Eq. \ref{BaeTauCooling})   obtained by \cite{Bae2016} at the midplane at the planet location. \par
In Fig.\ref{Fig:BaeFig6} we show the midplane density in the $(r,\varphi)$ plane when the planet has just reached its final mass. Multiple spiral arms are clearly shown (similar to Fig.6 of \cite{Bae2016}).

\begin{figure}
\centering
\includegraphics[width=\hsize]
   {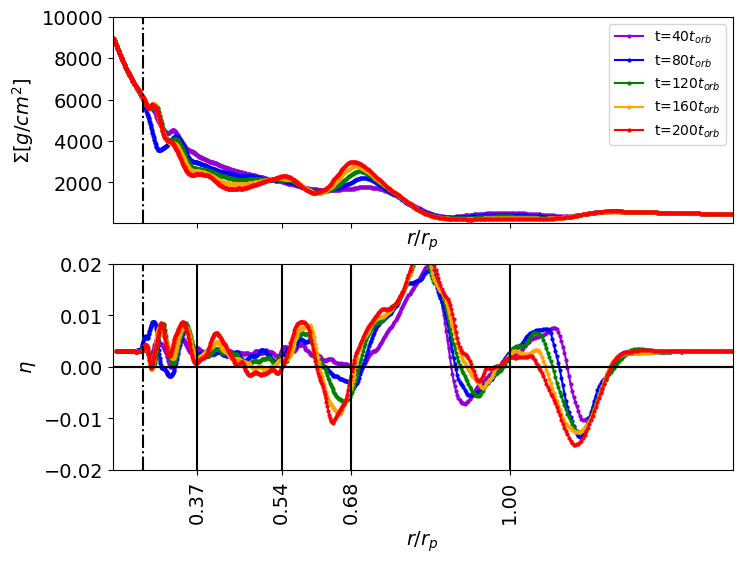}
   \caption{{\bf Top panel}: azimuthally averaged surface density profile for simulation HD$_{Bae}$ . {\bf Bottom panel} azimuthally averaged $\eta$ parameter. The planet is kept on a fixed circular orbit at $r/r_p=1$ and the ticks on the $x$-axis  (as well as the vertical black lines) indicate the  values of $\eta \sim 0$ with positive slope. The horizontal black line corresponds to $\eta =0$. The damping region extends radially from $r_{min}/r_0$ to the dot dashed line. }
      \label{Fig:BaeEta} 
\end{figure}

\begin{figure*}
\centering
\includegraphics[width=\textwidth]
   {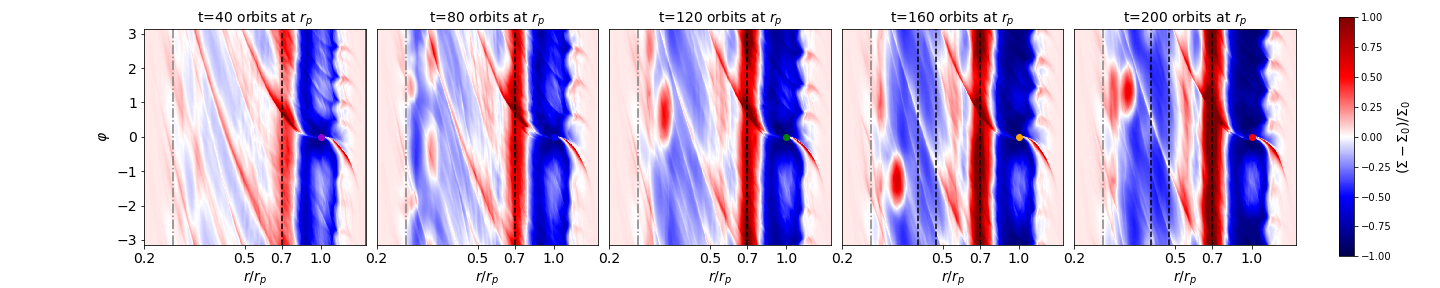}
      \caption{ Perturbed surface density of the HD$_{Bae}$ simulation at the same snapshots represented in Fig.\ref{Fig:BaeEta}. The dashed lines indicate the radial position of the pressure bumps, i.e. the values of r for which $\eta=0$ in Fig.\ref{Fig:BaeEta}. The dot-dashed line indicates the boundary of the damping region.}
         \label{Fig:BaeJupdelta} 
\end{figure*}

We consider  the same snapshots of Fig.5 of \citet{Bae2016} and show in  Fig.\ref{Fig:BaeEta} the azimuthally averaged surface density profile (top panel) and the
$\eta$ profile (bottom panel). Additionally to the first bump at the gap's inner edge at about $r/r_p=0.7$ another density maximum associated to a pressure bump is located at  $r/r_p=0.54$ and a third maximum at about 0.4 where $\eta$ gets close to zero.
Two additional pressure maxima appear inside $r/r_p=0.37$ but they do not correspond to rings.
Precisely, in  Fig.\ref{Fig:BaeJupdelta} the perturbed surface density plotted at the same   snapshots of Fig.\ref{Fig:BaeEta} clearly shows that the averaged maxima correspond to  rings extending over the full disk in azimuth excluding the two innermost maxima which correspond to vortices.

 In conclusion   a Jupiter mass planet can form multiple rings and gaps inside its orbit in a non viscous disk. Such rings are effective pressure bumps and can \red {potentially} stop dust from drifting towards the star. \red{ We stress the term  "potentially",  because $\eta$ is computed from the azimuthally averaged pressure and the actual response of dust to non-axisymmetric features (e.g. spiral waves) needs to be investigated in a future work}.

\section{Hydrodynamical case with complete treatment of radiative transport}
\label{Sec:rad}
The aim of this section is to study the propagation of density waves with a  self-consistent treatment  of radiation transport within the FLD approximation.\par
With respect to the study presented in \cite{Bae2016} and revisited in Section\ref{Sec:bae} we also consider a non zero viscosity. 
Actually, disks have low but non zero bulk viscosity as evidenced by  observations   \citep{pinte_dust_2016,villenave_highly_2022} and  
supported by mechanisms like the vertical shear instability \citep{Nelson2013} that provide small viscosity in dead zones of the disks where viscosity is not sustained by magneto-rotational instability.
The value of the alpha parameter $\alpha_{\nu}$ is expected to be lower than $10^{-3}$ and possibly  takes values of the order of
$10^{-4}$. We will consider in the following $\alpha_{\nu}=10^{-4}$.
\par

\begin{figure}
\centering
\includegraphics[width=\hsize]
   {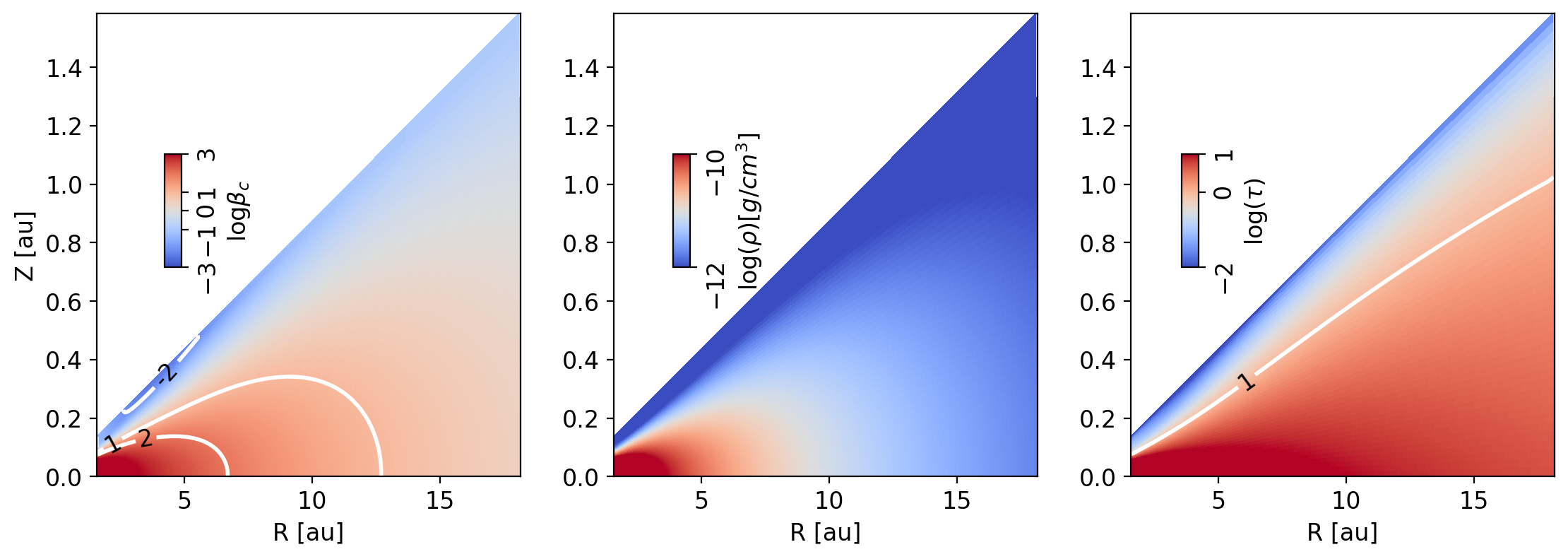}
      \caption{Disk cooling time, density and opacity in the $(R,Z)\equiv (r\sin(\theta),r\cos(\theta))$ plane of the disk in thermal equilibrium
      before the insertion of the planet. }
         \label{Fig:tau0} 
\end{figure}

\begin{figure}
\centering
\includegraphics[width=\hsize]
   {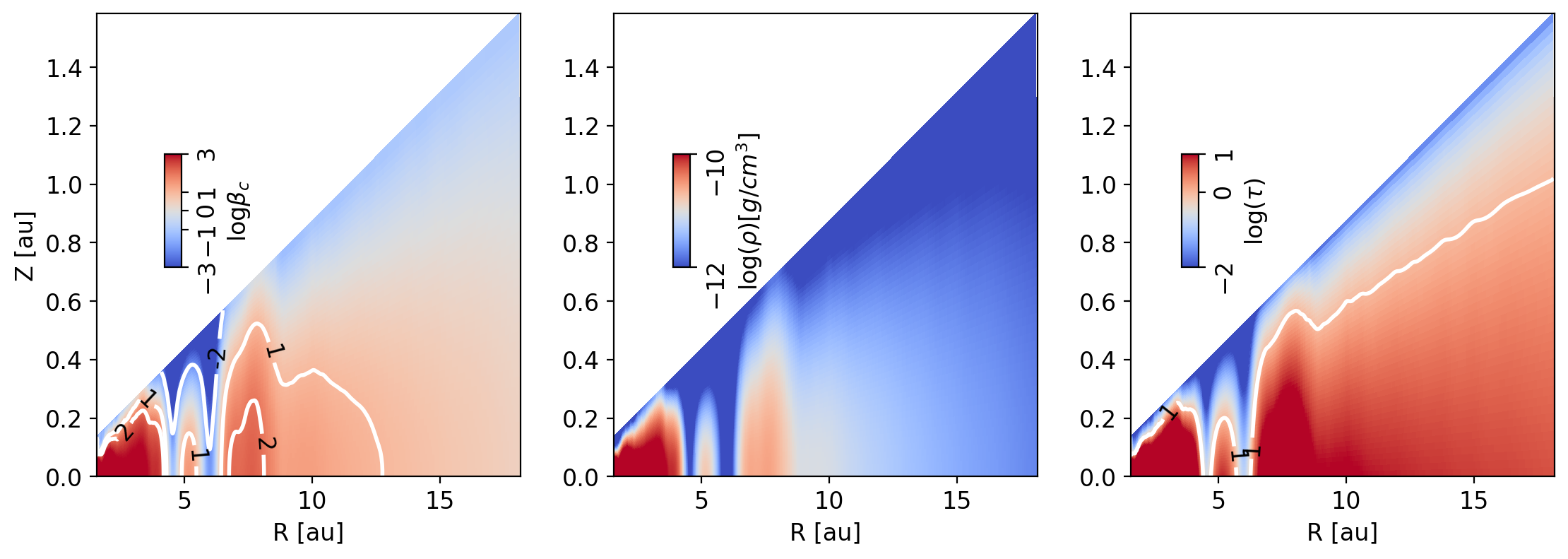}
      \caption{Same as Fig.\ref{Fig:tau0} for azimuthally averaged quantities for the three dimensional simulation  HD$_{rad}$J with embedded a Jupiter mass planet. The snapshot corresponds to  the end of the integration at 150 orbital periods at the planet location ($5.2 au$).}
         \label{Fig:tau150} 
\end{figure}

Before inserting the planet  we run a 2D axisymmetric ($r,\theta)$ simulation until the disk reaches thermal equilibrium.
We notice that the main source of heating in a low viscosity disk is the stellar heating. In order to recover the temperature profile of a stellar irradiated disk we run a 2D ($r,\theta)$ simulation including the heating from a Solar type star  and use the temperature values of the resulting equilibrium disk as boundary  for our  three dimensional disk at the disk's surface (as described in \cite{2024A&A...690A.183L}).
The equilibrium disk is flared  (with flaring index 2/7) and has an aspect ratio of  $0.03$ at $r_p$. 
The density slope $\alpha_{\Sigma}$ (see table \ref{table:tab1}) is chosen to have a disk with constant star accretion rate \citep{Legaetal2015}.  The mass flow carried by the disk is set to be $\dot M \sim 10^{-9}M_{\odot} / yr$, about one order of magnitude smaller than  the typical accretion rate  observed in young stars \citep{1998ApJ...495..385H,2016A&A...591L...3M}. We will recover typical star accretion rates by including non ideal MHD effects in Section \ref{Sec:MHD}.
\par
Fig. \ref{Fig:tau0} shows the values of the equilibrium disk cooling time, density and opacity in the $(R,Z)$ plane. 
 When looking at values of $\beta_c$ in Fig.\ref{Fig:tau0} left panel, the disk can be divided in two main regions by the iso-contour line $\log \beta_c =1$. Precisely,
inside the contour  $\log \beta_c =1$ (Fig.\ref{Fig:tau0}, left panel) the cooling time values
  correspond to the regime of density wave propagation  called "adiabatic" in \cite{Miranda2020} (see their Fig.4). When inserting a planet in this region we expect a gap at the planetary orbit and multiple rings and gaps in the inner region.  Near the  $\log \beta_c =1$ contour, density waves are expected to be radiatively damped (see \cite{Ziampras2023}). Finally, for $\log \beta_c \ll 1$  the disk is in the locally isothermal regime. This regime is found near the the disk surface, far from the planet's forming region. We remark that the $\beta_c$ values will change with time according to the perturbation introduced by the planet and  wave propagation regimes can be modified accordingly.

\subsection{Jupiter mass planet}
 Starting from the $(R,Z)$ disk  in thermal  equilibrium, we expand it in azimuth  over the interval $[-\pi,\pi]$ and restart our simulation with an embedded planet. 
 We run the HD$_{rad}$J simulation over 150 orbits at the planet location and show the azimuthally averaged values of $\beta_c$, of the volume density and the optical thickness at the end of the integration on Fig.\ref{Fig:tau150}. The planet has carved a gap and deeply modified the disk structure. However, the values of  $\beta_c $ inside the planetary orbits are still compatible with the "adiabatic" regime of wave propagation so that we expect to find multiple gap and rings in the inner disk region ($r <4 au$).
 \par
\begin{figure*}
\centering
\includegraphics[width=\textwidth]
   {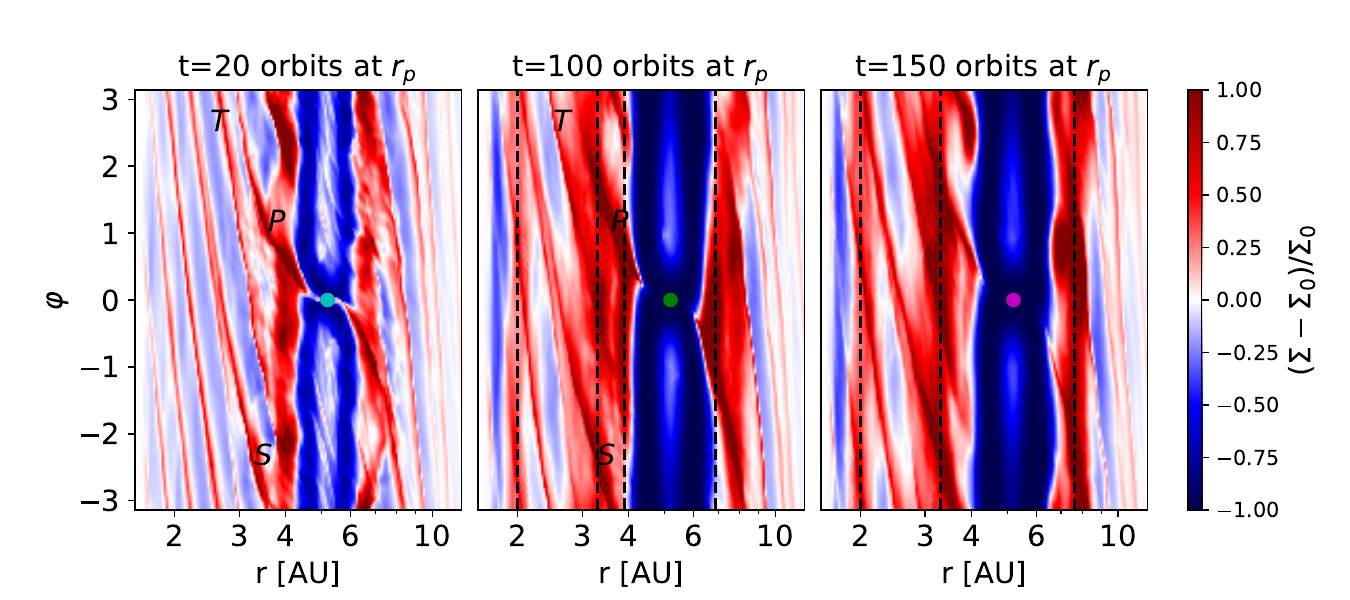}
      \caption{Perturbed surface density of the HD$_{rad}$J simulation at different snapshot indicated in orbital period at the planet location. The letters in the left and central panels indicate the primary (P), secondary (S) and tertiary (T) spiral arms. The dashed lines indicate the position the radial position of the pressure bumps, i.e. the values of r for which $\eta=0$, with positive slope in Fig.\ref{Fig:JupSE}.  }
         \label{Fig:Jupdelta} 
\end{figure*}
\begin{figure}
\centering
\includegraphics[width=\hsize]
   {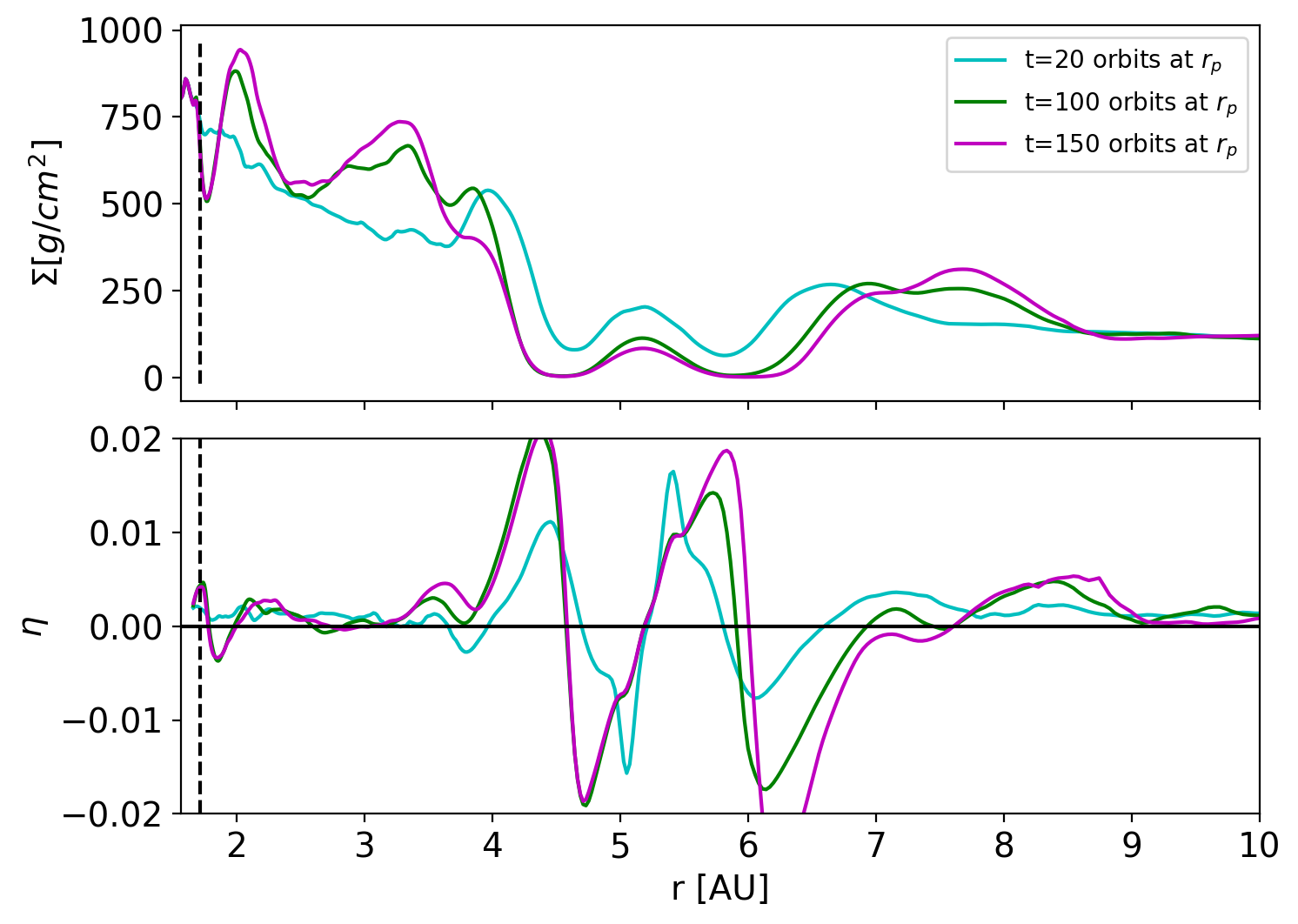}
      \caption{{\bf Top panel}: azimuthally averaged surface density profile for simulation $HD_{rad}J$ at the same snapshot represented in Fig.\ref{Fig:Jupdelta}. {\bf Bottom panel} azimuthally averaged $\eta$ values. The dashed line indicates the boundary of the damping region. }
         \label{Fig:JupSE} 
\end{figure}

In Fig. \ref{Fig:Jupdelta} we show the perturbed surface density of the HD$_{rad}$J simulation at different snapshots indicated in orbital period at the planet location: in the left panel, $t=20$ orbits, we clearly see the gap at the planet's orbit and a primary ring of gas at the inner and outer gap edges (with density maxima respectively at $r \sim 4$ and $\sim 6.7 au$) as well as the primary spiral arm.  A secondary and a tertiary  arms are clearly present inside the planetary orbit launched respectively at  $(r,\varphi) \sim (3.8,-\pi)$ and $(r,\varphi) \sim (3.8, -1.8)$ and propagating in the inner disk. 
The presence of  multiple spiral arms is a necessary condition to have multiple gaps \citep{2019ApJ...875...37M,2020ApJ...892...65M}. As for the primary gap at the planet's orbit, multiple gaps form when density waves  steepen into shocks depositing angular momentum into the gas. A second and a third ring  at respectively $r \sim 3.3$ and $2 au$ are clearly formed at $t=100$ (Fig.\ref{Fig:Jupdelta}, second panel and Fig. \ref{Fig:JupSE}, top panel). \par
Over the total integration time of $150$ orbits the primary gap is still evolving and rings of density maxima
slightly move inward and outward as indicated by the vertical dotted lines in Fig\ref{Fig:Jupdelta}.

The density contrast of the innermost ring increases with time while the second ring at    $r \sim 3.3$
partially merges with the primary inner ring as the former eventually  moves inward.

From the computation of $\eta$  (Fig. \ref{Fig:JupSE}, bottom panel) we can say that there is a pressure bump at the  innermost ring at about $2 au$  so that  a reservoir of dust should be trapped there, while the primary and secondary rings are marginally separated at $t=150$ orbits. The pressure bump at about $3.3 au$ should stop dust from drifting inward and make a  distinct reservoir with respect to the one centered at  $2 au$.
\par
Three dimensional fully radiative simulations require about a factor 10 longer computational time with respect to isothermal or adiabatic simulations with controlled cooling \footnote{The diffusion part of the energy equation is solved implicitly and this often requires a large number of iterations for the numerical scheme to reach convergence.}. \par The important result here is  that distinct dust reservoirs can be formed in a realistic scenario where thermal processes are computed in a self consistent way.

  \subsection{Comparison with various EoS}
  In this subsection we consider simulations with an embedded  Jupiter-mass planet like in the  HD$_{rad}$J case but using a different EoS
  in order to  compare results obtained in HD$_{rad}$J with different wave damping regimes  (see \cite{Miranda2020,Ziampras2023}). In the following  we consider disks with respectively  an isothermal EoS (HD$_{iso}$),  an adiabatic EoS with temperature relaxation to the initial value as in Eq.\ref{eq:EdotRelax}  (HD$_{\beta_c}$ with  $\beta_c=1$ and $\beta_c=100$) and an adiabatic case without thermal relaxation (HD$_{adia}$).
  We have integrated the new simulations on longer times with respect to HD$_{rad}$J taking  advantage of shorter running times \footnote{About 2.5 hours for 10 orbits at the planet location on 1 GPU AMD MI250 to be compared with 2.5 hours for 1 orbit on 96 cores on AMD Genoa EPYC 9654} in order to  check the persistence of rings and gap structures.
  Moreover, the disk domain is  extended  further in  to investigate the possible formation of rings  inside the one at $2 au$ obtained in  the case HD$_{rad}$J and also to confirm that this innermost ring at  $2 au$  is not affected by the close inner disk boundary (which is at $~1.5 au$ in HD$_{rad}$J ).\par
\begin{figure*}
\centering
\includegraphics[width=\textwidth]{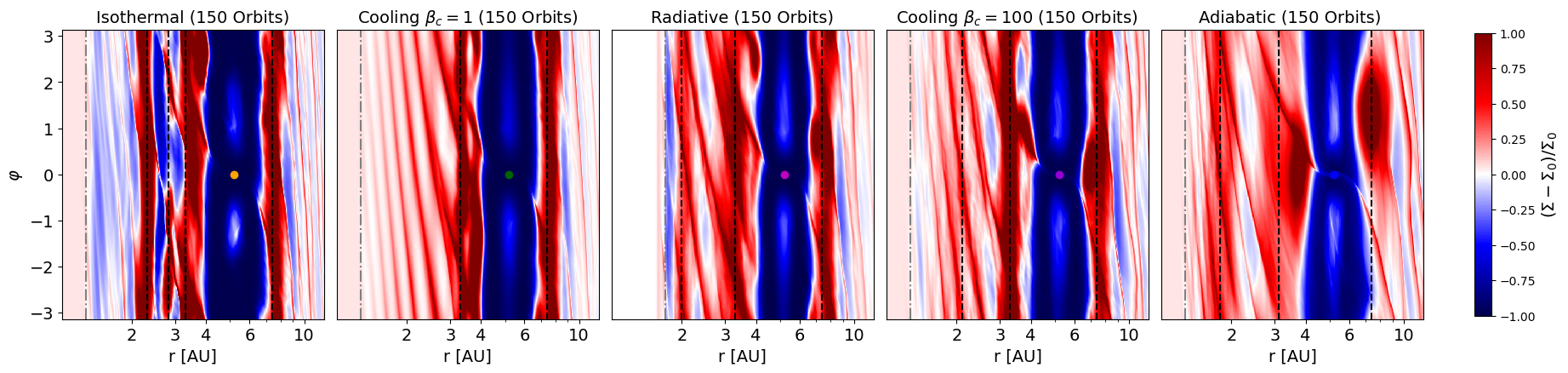}
\includegraphics[width=\textwidth]{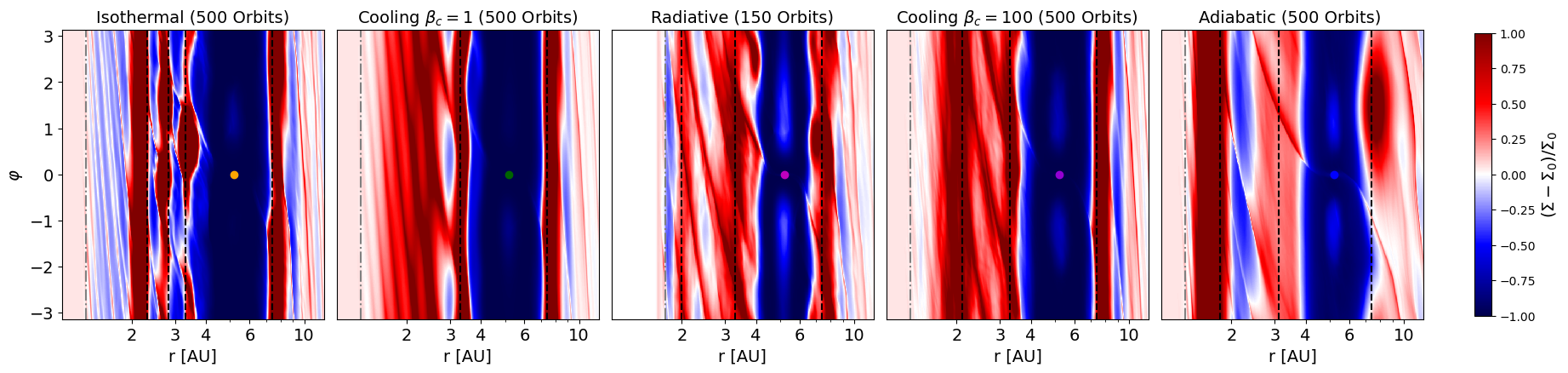}
      \caption{ Perturbed surface density at $t=150$ orbits
      at the planet location (top panels) and  at $t=500$  (bottom panels) orbits. From left to right the simulation sets are: HD$_{iso}$, HD$_{\beta_c}$ with  $\beta_c=1$, HD$_{rad}$ , HD$_{\beta_c}$ with  $\beta_c=100$, HD$_{adia}$.  We observe multiple rings formation which depend on  the EoS. Rings are persistent on the longer integration time. For the radiative case we report, on the bottom panel, the last snapshot, i.e. the one at 150 orbits.}
         \label{Fig:IJupdelta} 
\end{figure*}
\begin{figure}
  \centering
\includegraphics[width=\hsize]{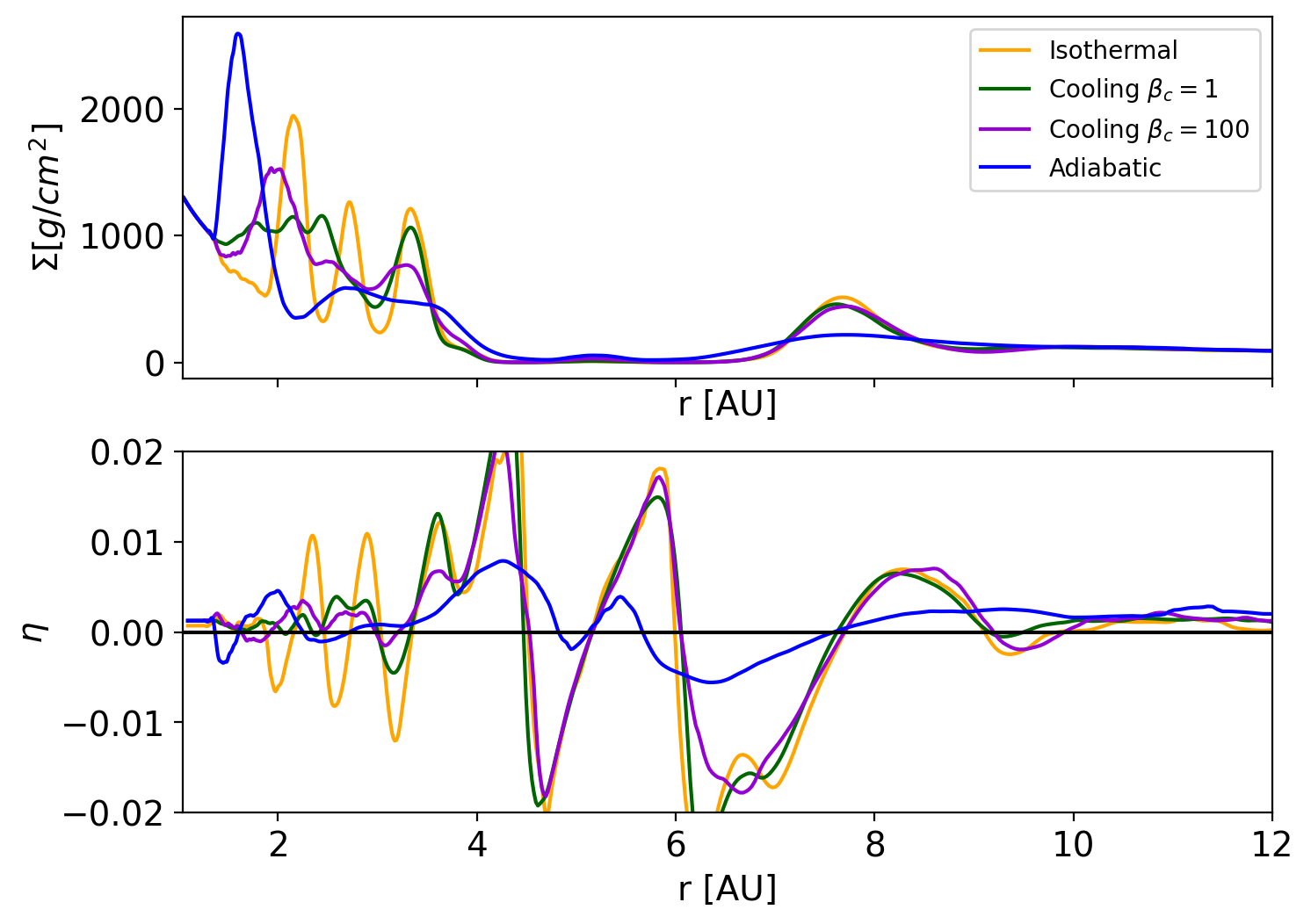}
      \caption{Azimuthally averaged surface density profile ({\bf Top panel}) and the  $\eta$ values ( {\bf Bottom panel}) for simulation sets HD$_{iso}$, HD$_{\beta_c}$, HD$_{adia}$ at $t=500$ orbits at the planet location.}
         \label{Fig:MultiJupSE} 
\end{figure}

  We show in Fig.\ref{Fig:IJupdelta} the perturbed surface density at $t=150$ orbits at the planet location (top panels) and at $t=500$ orbits (bottom panels)  for all these cases and,
  for comparison, we insert, in between  the two HD$_{\beta_c}$ cases, the results obtained before for the  HD$_{rad}$J case at $t=150$ orbits.
  As for the two dimensional case \citep{Ziampras2020}, the locally isothermal disk is the most efficient in forming multiple rings and gaps  : three in our disk which are also very contrasted (see Fig.\ref{Fig:MultiJupSE}, top panel) and are effective dust barriers based on the radial profile  of $\eta$  (Fig. \ref{Fig:MultiJupSE}, bottom panel). We  do observe spiral arms extending further in but no additional ring  is observed inside  $r=2 au$. \par
  At $t=500$ orbits the rings  appear more contrasted with respect to  $t=150$ orbits, which is a good indication for their persistence.
  In all the cases an innermost ring is formed closed to $r=2 au$ except in the case with $\beta_c=1$ which, according to the results obtained for two dimensional disks, is the case in which wave damping appears to be the  most efficient (when no radiative effects are taken into account see \cite{Ziampras2023}).
  \par We notice the similarity between the fully radiative case and simulation HD$_{\beta_c}$ with $\beta_c=100$, i.e with a cooling time consistent with typical cooling times inside the Jupiter orbit  in Fig.\ref{Fig:tau150}. 
  Also, we can say that the ring at $r=2 au$
  is not affected by the close radial boundary since a very similar structure is obtained in disks extending further in.

\section{MHD simulation with prescribed cooling}
\label{Sec:MHD}
In the previous section we have shown that the opening of secondary gaps by a  Jupiter mass planet is observed in low viscosity three dimensional disks with self-consistent treatment of the diffusion of heat.
\el {We will consider the formation history of Jupiter by analyzing smaller masses  in Section \ref{Sec:MinMass}. 
In this section we notice that }
low viscosity disks do transport gas towards the star at a very low rate with respect to  typical accretion rates  observed in young stars \citep{1998ApJ...495..385H,2016A&A...591L...3M}.
Instead, in magnetized disks  embedded in  warm fully ionized corona,   gas transport at disk surface layers  (at about 2-3 disk scale height) can provide the accretion flow onto the star that we did not have in our simulation HD$_{rad}$J. Actually, gas can flow inward because of  angular momentum removal by disk winds \citep{Bethune2017,Lesur2021}.\par   
In the simulations presented here we do not solve the radiation field but we prescribe a warm corona through a steep vertical increase  of the disk midplane temperature from 3 to 5 pressure scale height.
Precisely, \el { similar to \cite{Bethune2017}}, the disk is initialized with a temperature ratio between the corona and the disk $T_{cd} = 6 $ as follows:
\begin{equation}
    \label{Tcd}
T(\tilde \theta) = 1 + (T_{cd}-1)\left(\frac{1}{1+\exp[-6(|\tilde \theta| - 4h)/h]}\right)^2
\end{equation}
where $\tilde \theta \equiv \pi/2-\theta$ is the latitude, $h$ is the disk aspect ratio (constant as our MHD disks are considered non-flared, see Appenxix \ref{App:MHDInit}).

To this aim we assume for the sound speed a function
\begin{equation}
    c_s(r,\theta)= \sqrt{\frac{GM_{*}}{r}h^2T(\tilde \theta)}
\end{equation}
Considering the equations for vertical and rotational equilibrium we provide in  Appendix \ref{App:MHDInit}
the initial volume density  and the initial azimuthal velocity that we use in our MHD simulations.
\par
After disk initialisation we increase $T_{cd}$ to 10 in the time interval $[0:30]$ (code units).\par
The initial magnetic field is vertical $\bf {B} = B_z$.
Although there is a large uncertainty about the topology of the  magnetic field of protoplanetary disks, their intensity is known to be weak \citep{Lesur2021,Lesur2021b} in the sense that the midplane ratio of thermal over magnetic pressure: $\beta \equiv 2\mu_0P/B_z^2 \gg 1$. Therefore we  consider disks with initial midplane values $\beta = 10^{6}, 10^{5} , 5\times 10^{3}$
and compute the radial profile of the magnetic field at midplane from $\beta$ and from the midplane pressure: $P=\rho(r,0)c_s(0)^2$. 
\par
We do not solve the energy equation with the FLD 
approach (Eq. \ref{eq:Edot2Temp}) as in simulation HD$_{rad}J$ since this  would require extremely long integration times for three dimensional MHD simulations. \par However,
we have shown in Fig.~\ref{Fig:IJupdelta} that, when relaxing the internal energy  to the initial temperature profile  according to Eq.\ref{eq:EdotRelax}  
on a  cooling time $\beta_c=100$ we obtain results very similar to the fully radiative case. 
Therefore, we consider Eq.\ref{eq:EdotRelax}   in our MHD simulations  with cooling time: $\beta_c = 100$.
\par
\begin{figure}
\centering
\includegraphics[width=\hsize]{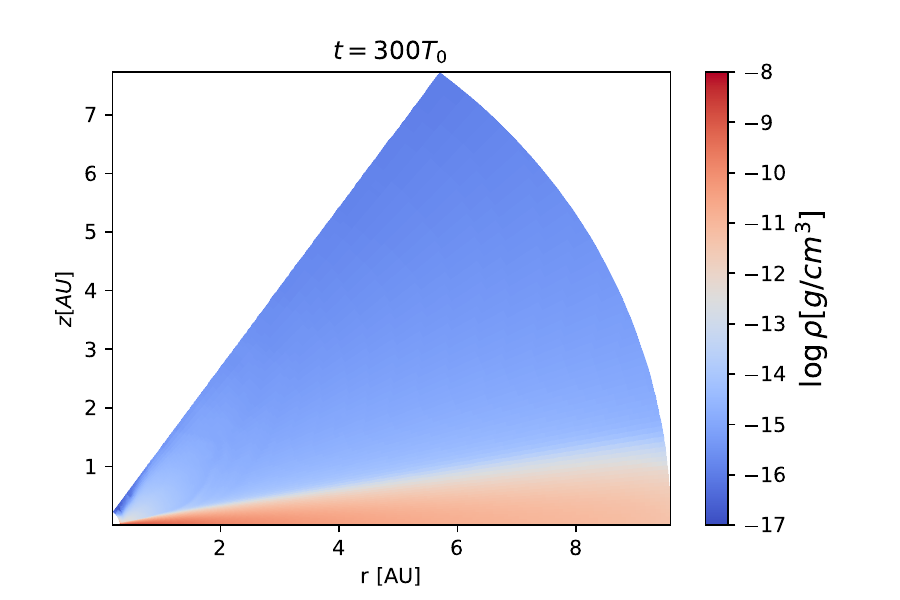}
\includegraphics[width=\hsize]{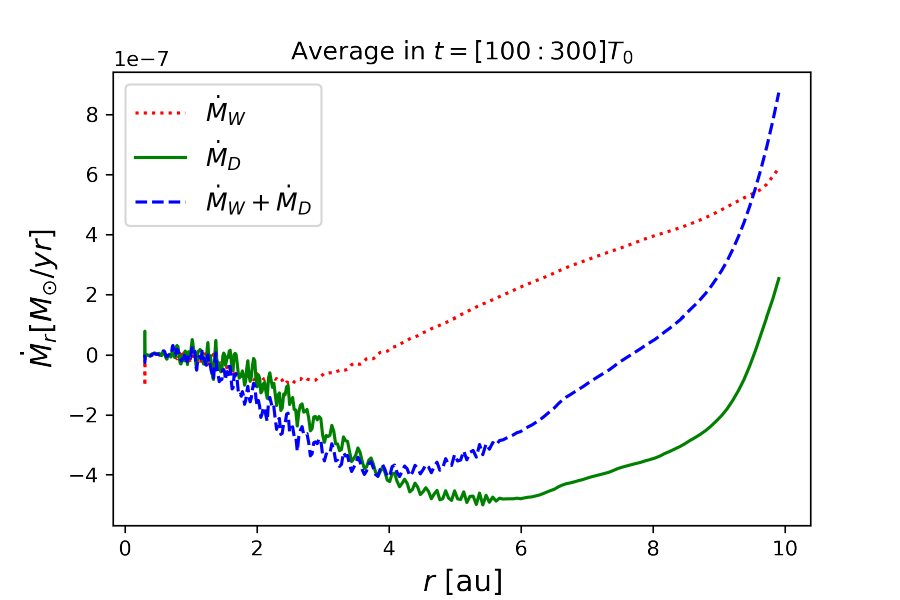}
\caption{{\bf Top panel}  Density distribution in the $(r,Z)$ plane obtained for our simulation 
 MHD$_{\beta 5e3}$ with $\Lambda _A 1$ at $t=300 T_0$. {\bf Bottom panel} Computation of $\dot M(r,\theta)$ (Eq.\ref{mdot}),  averaged over the time interval $[100:300]T_0$.}
\label{RadialMdot}
\end{figure}

Before inserting the planet we run two dimensional $(r,z)$ simulations for a few hundreds   orbits at $r=1$ au.
At $t=300 T_0$ ($T_0 = 2\pi$~yr) we show (Fig.\ref{RadialMdot},top panel)   the density distribution on the $(r,Z)$
plane for the case $\beta = 5\times 10^{3}$ and $\Lambda_A =1$ (fiducial simulation). It appears clearly from  (Fig.\ref{RadialMdot},top panel) that, in the regime of Ohmic diffusion dominating over the Ambipolar one, we do not see a spontaneous generation, i.e. in absence of the planet the density distribution does not show rings and gaps (see for example \cite{Bethune2017}).  Therefore, when inserting a planet, gap and ring formation will be a consequence of the perturbation induced by the planet itself.
We also compute the mass flux as a function of  $r$ as:
\begin{equation}
\label{mdot}
\dot M =2 (\dot M_d +\dot M_w)
\end{equation}
where the factor 2 takes into account the full disk.
\el {The quantities $\dot M_d$ and $\dot M_w$ are the mass flux of respectively the disk region up to $2H$ and the corona
from $2H$ to the disk surface.
These quantities are computed from:
\begin{equation}
\label{mdotD}
\dot M_d = \int _0^{2\pi}\int _{\theta_{2H}}^{\pi/2}\rho v_r r^2sin(\theta)d\theta d\varphi
\end{equation}
with $\theta_{2H}$ the colatitude at 2 disk scale height
and:
\begin{equation}
\label{mdotC}
\dot M_w = \int _0^{2\pi}\int _{\theta_{0}}^{\theta_{2H}}\rho v_r r^2sin(\theta)d\theta d\varphi
\end{equation}
with $\theta(0)$ the value of the colatitude at the disk surface. }
In the bottom panel of Fig.\ref{RadialMdot} we plot this quantity 
 averaged in time from $100T_0$ to $300T_0$ for our fiducial simulation.  Negative values  correspond to flux directed towards the star, while positive ones correspond to gas outflow. The corona appears to be dominated by disk out flowing wind while the disk
 accretes  with values of   $\dot M \sim 10^{-7}M_{\odot} / y$,  similar to  typical accretion rate  observed in young stars \citep{1998ApJ...495..385H,2016A&A...591L...3M}. 
 We provide in Appendix \ref{App:MHDDiag} an analysis of the mass flux from angular momentum conservation arguments. 

\begin{figure*}
\centering
\includegraphics[width=\textwidth]{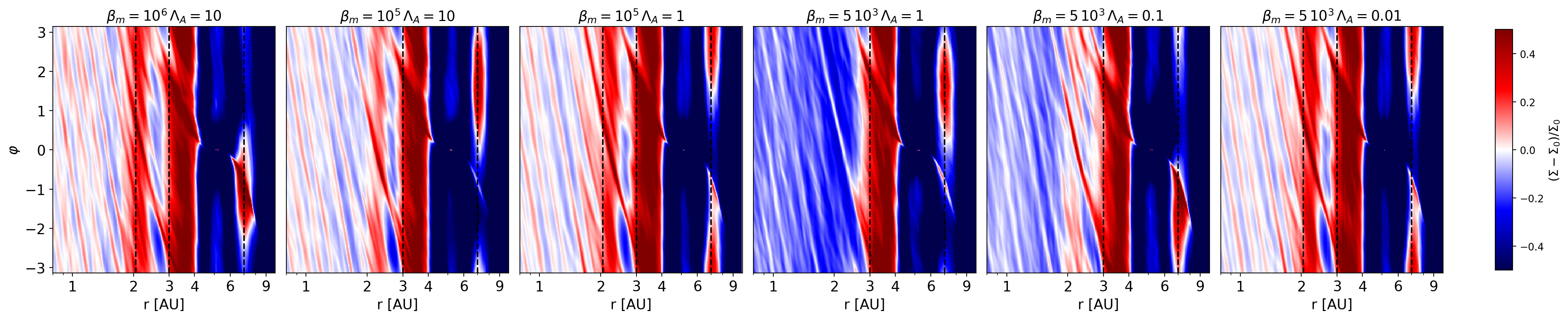}
      \caption{Perturbed surface density of the MHD simulations. From left to right we increase the strength of the magnetic field and for a given value of the magnetic field we may also increase the Ambipolar coefficient value.  For the lowest value of the magnetic field (left panel) we see the formation of a secondary ring at $r=2$ au with our lower value of  Ambipolar diffusion ($\Lambda_A=10$),
      so that we do not run additional simulations. 
      In the case $\beta_m= 10^5$  (second panel and third panels) we observe a secondary ring for $\Lambda_A=10$ , however the formation of a secondary well contrasted ring requires a larger  Ambipolar diffusion strength ($\Lambda_A=1$, third panel). Finally, for $\beta_m= 5\times 10^3$ there is no secondary ring formed for $\Lambda_A=10$ (not shown) nor for $\Lambda_A=1$ (fourth panel). We observe a poorly contrasted secondary ring for $\Lambda_A=0.1$  (fifth panel) and we finally observe a well contrasted secondary ring for
      $\Lambda_A=0.01$ (rightmost panel).}
         \label{Fig:MHDSigma} 
\end{figure*}

We let the planet growing for $t=10$ orbital periods at the planet location and continue our simulation over $100$ orbits.
In Fig.\ref{Fig:MHDSigma}  we show the perturbed surface density of our MHD runs at 
$t=100$ orbits at $r_p=5.2$ au.

 \begin{figure}
 \centering
 \includegraphics[width=\hsize]
   {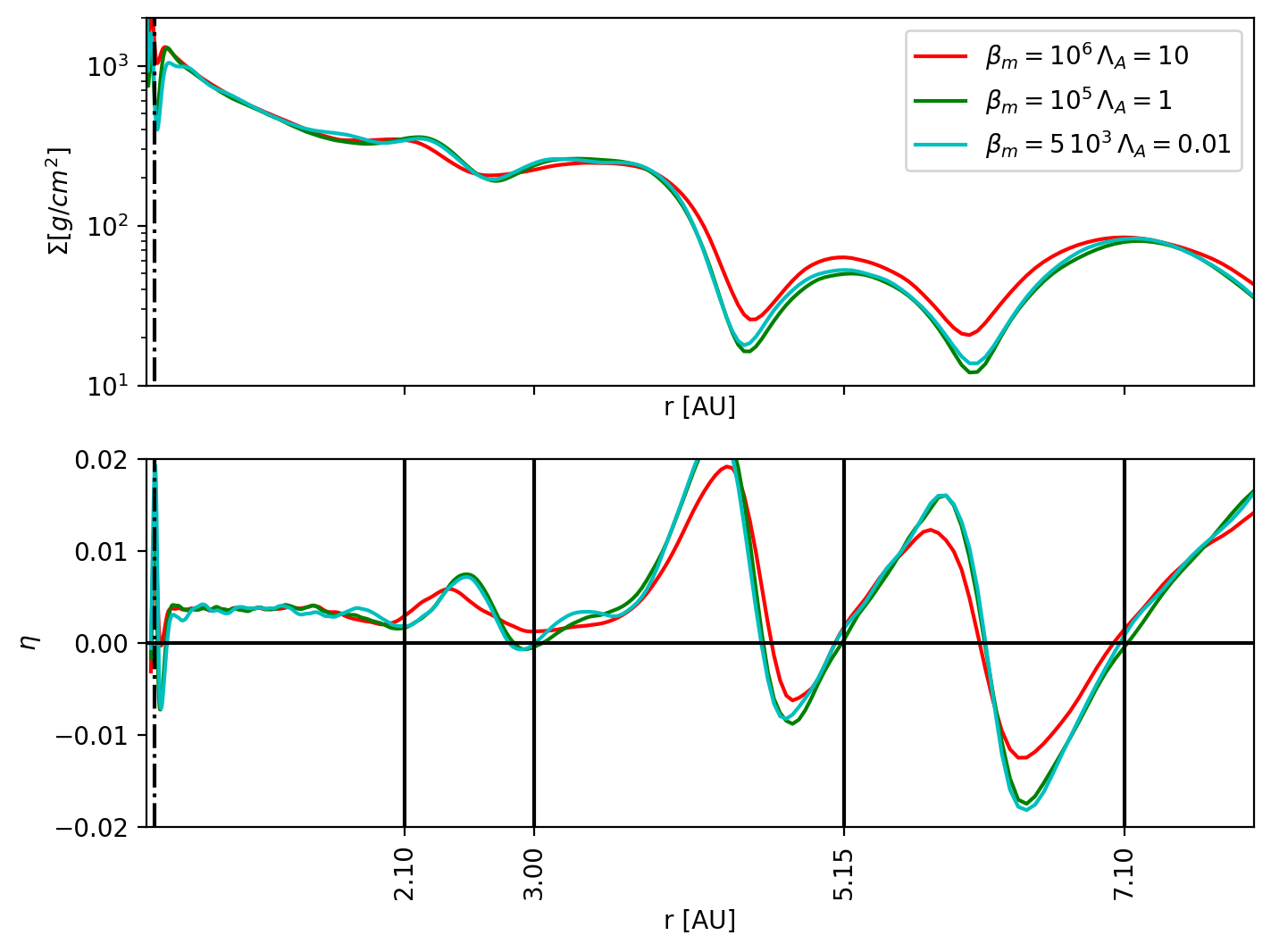}
      \caption{Azimuthally averaged surface density profile (top) and azimuthally averaged $\eta$ values (bottom)  for MHD simulations  presented in Fig.\ref{Fig:MHDSigma} in which a well contrasted secondary ring is formed. The planet is kept on a fixed circular orbit at $r_p=5.2 \, au$ and the ticks on the $x$-axis (as well as the vertical black lines) indicate the  values of $\eta \sim 0$ with positive slope. The horizontal black line corresponds to $\eta =0$.  The damping region extends radially from $r_{min}$ to the dotted  line.}
         \label{Fig:MHDSE} 
 \end{figure}

 From left to right we increase the strength of the magnetic field and for a given value of the magnetic field we may also increase the Ambipolar diffusivity depending on the results obtained for multiple rings formation. For the lowest value of the magnetic field ($\beta =10^{6}$, left panel) we see the formation of a secondary ring t about $r=2 \, au$. We have considered in this case $\Lambda_A=10$, which correspond to a very low effect of the non ideal ambipolar diffusion effect. Keeping this same value of $\Lambda_A$, and increasing the strength of the magnetic field ($\beta =10^{5}$, second panel), we barely observe  the formation of a secondary ring.  Instead, by increasing the ambipolar coefficient ($\Lambda_A=1$, third panel)
 we observe a significant enhancement of the secondary ring. 
\el{ By further increasing the magnetic field ($\beta=5\times 10^{3}$, right panels), we observe that the inner planet-induced spiral arms can be damped, in which case no secondary ring ever forms  at $\Lambda_A=10$ (not shown) and $\Lambda_A=1$ (fourth panel), while a weakly contrasted secondary ring appears for $\Lambda_A=0.1$ (fifth panel). 
 Finally, a well contrasted secondary ring appears in the rightmost panel of Fig. \ref{Fig:MHDSigma} for  $\Lambda_A=0.01$ .}
 \par
\el{  For the largest value of $\beta$, the magnetic field is immaterial and the formation of secondary rings closely resembles that occurring in purely hydrodynamical disks, regardless of the Ohmic and ambipolar diffusion coefficients. As the magnetic field increases (for smaller values of $\beta$), MHD effects come into play and the magnitude of ambipolar diffusion is found to have a strong impact on the existence of secondary rings. }
 \par
 This effect is similar to the one observed in the  formation of zonal flow \citep{Bethune2017,Riols2020} in disk's regions where Ambipolar diffusion is the dominant non ideal effect.
 \el {Precisely, spontaneous accumulations of gas  are observed with anti-correlated concentrations  of the vertical component of the magnetic field $B_z$. It has been shown \citep{Bethune2017}  that  ambipolar diffusion is  responsible for the accumulation of $B_z$. 
 Therefore, if one or multiple gaps forms in a non-spontaneous way, for example because a massive planet perturbs the gas, then $B_z$  will accumulate in the gaps  \citep{Wafflard2023}  and enforce gas accumulation  in rings provided that ambipolar diffusion is strong enough}.
 \par 
 
 In Fig.\ref{Fig:MHDSE} we show the azimuthally averaged surface density  (top panel) and the  $\eta$ values (bottom panel) as a function of $r$ at the same snapshot of Fig.\ref{Fig:MHDSigma}. 
 The plots correspond to the simulations presented in Fig.\ref{Fig:MHDSigma} for which  a clearly contrasted secondary ring is formed: i.e. for $\beta = 10^6$ with $\Lambda_A =10$, for  $\beta = 10^5$ with $\Lambda_A =1$ and for  $\beta = 5\times 10^3$ with $\Lambda_A =0.01$. \par
 The values of $\eta$ for both the primary and  secondary rings at $r~2$ au reach values close to zero although we do not observe a clear transition from negative to positive $\eta$ which is necessary for dust-trapping. A longer integration time would be  required to \red{ verify whether} dust-trapping \red{is possible}  in these  rings. Nevertheless,  we consider very interesting that  rings form for values of non-ideal parameters that are plausible in  the disk region inside the Jupiter orbit.

 \section{Minimum planet mass forming dust-trapping rings}
\label{Sec:MinMass}
It is well known \citep{Lambrechts2014,2018A&A...612A..30B}   that  when a giant planet  core reaches few tens of Earth masses the so-called pebble isolation mass is reached and dust of suited Stokes numbers is blocked at the outer gap's edge opened by the core. Moreover, in low viscosity disks even cores of less than 20 Earth masses can open small gaps lowering the limit of the pebble isolation mass \citep[see][]{2018A&A...612A..30B}. \el { On  the other hand}, the dust inside the planet's orbit  drift towards the star and is eventually stopped at the inner (close to 1 au) pressure bump  due do gas removal by magnetized winds \citep{Suzuki2010,Ogihara2018}. \par
\el {We have shown in this paper that a Jupiter mass planet has the ability to form multiple rings, according to disk thermal properties and to non ideal magnetic diffusivities. However, such rings can be effective dust's reservoirs  if they form early on in the Jupiter formation history, possibly when Jupiter core has reached or slightly overcome   the pebble isolation mass.}\par
Therefore, in this section we run simulations for a planet of Saturn mass (simulation HD$_{rad}$S) and for a super Earth of 30 Earth masses (simulation HD$_{rad}$30) in order to check their ability to form multiple rings.
\par
\begin{figure}
\centering
\includegraphics[width=\hsize]
   {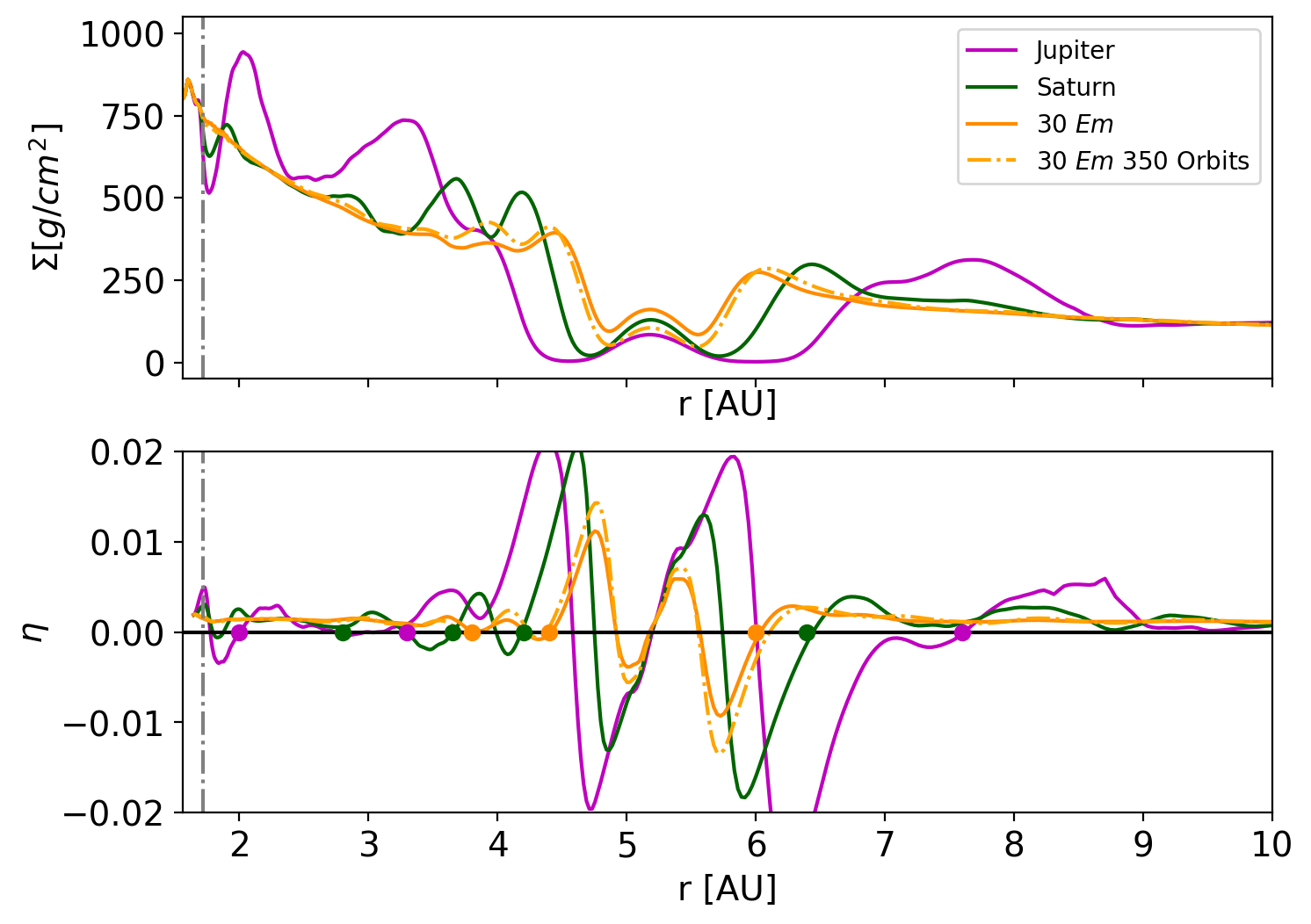}
      \caption{{\bf Top panel}: azimuthally averaged surface density profile for simulations $H_{rad}30$, $H_{rad}S$ and
       $H_{rad}J$ at 150 orbital periods at the planet location. {\bf Bottom panel} azimuthally averaged $\eta$ values. \el{ Color markers at $\eta = 0$  with positive slope identify the position of the pressure bumps on the top panel.}}
         \label{Fig:MassSE} 
\end{figure}

The super Earth opens a gap and forms a secondary ring  \el {(at $\sim 3.8$ au)} which increases in density contrast on long integration times (see Fig.~\ref{Fig:MassSE}). A third density maximum appears which, however, is not associated to a local pressure maximum since it has $\eta>0$. Instead,  3 rings appear for Saturn at 150 orbits and 2 for Jupiter as previously discussed. The values of $r$ for which $\eta=0$ move inward when increasing the planet's mass.

From Fig.\ref{Fig:MassSE}, we can propose the following evolutionary picture. When a planet reaches the pebble isolation mass it starts trapping dust at two locations near the inner edge of its gap. As the planet grows in mass, the locations of these two rings shift inwards, together with their dust. Near the mass of Saturn a third ring appears near 3 au, trapping there the dust that had not been trapped in the two other rings (unless the planet grows so slowly that all the untrapped dust has already drifted into the star).  As the planet grows further, the two initial rings merge and the last one moves from $\sim 3$ to $\sim 2$ au.

Obviously, this picture does not take into account that the planet is migrating at the same time. But, in the case of proto-Jupiter, its migration may have been stalled or even reversed by the appearance of a proto-Saturn \citep{2011Natur.475..206W}.

\section{Conclusion}
Cosmochemical observations provide evidence that mm-size dust remained trapped in the inner part of the disk for a timescale of at least a million years. Multiple trapping sites should have existed in order to explain the formation of four families of chondrites, which are chemically and isotopically distinct from each other. 

Thus, in this work we have explored the possibility that Jupiter formed  
\red {local maxima in the azimuthally averaged pressure } in the inner part of the protoplanetary disk (inwards of its orbit). 
\red{These averaged pressure bumps are potential sites of dust-trapping, but the actual response of dust to  the non-axisymmetric features that characterize these bumps remains to be investigated}. 

We have shown that in a low viscosity disk the opening of secondary gaps and density bumps is robust. It is observed in our 3D simulations also when we simulate self-consistently the diffusion of heat which, in the inner disk, gives a cooling timescale of about 100 orbital times. 

\el {When considering a strong magnetic field we observe that the inner planet-induced spiral arm can be damped and a secondary ring  forms only if ambipolar diffusion is strong enough.  Further work is warranted to assess the origin of this damping, which could arise from dissipation when the diffusion coefficients have adequate values, or from a removal of the angular momentum flux by the wind.}
\par
We remark that self-consistent disk ionization models show that the appropriate parameters for non-ideal MHD are consistent to those for which effective formation of secondary gaps and rings are observed in our simulations. 

A planet seems to be able to open secondary gaps and trap dust in rings when it reaches a mass close to the pebble isolation mass. Thus, as it cuts the flow of dust from the outer disk across its orbit, it also traps part of the dust in rings located inwards of its orbit. As the planet grows, these rings move away from the planet and can merge with each other, while a new ring can be formed farther away. 

Coupled with planet migration, these results provide a new view of a non-trivial evolution of dust in the inner solar system, which may be consistent with cosmochemical constraints on dust preservation and confinement on long timescales.

\begin{acknowledgements}
  AM is grateful for support from the ERC advanced grant HolyEarth N. 101019380. This work was granted access to the HPC resources of IDRIS and CINES under the allocation A0160407233 made by GENCI. F.~M. acknowledges support from UNAM's grant PAPIIT~107723,
UNAM's DGAPA PASPA program and the Laboratoire Lagrange at
Observatoire de la Côte d'Azur for hospitality during a one-year sabbatical stay. EL wish to thank Alain Miniussi for maintenance and re-factorisation of the code fargOCA. This work was supported by the
French government through the France 2030 investment plan
managed by the National Research Agency (ANR), as part
of the Initiative of Excellence Université Côte d’Azur under reference number ANR-15-IDEX-01.
\end{acknowledgements}
%
%
\bibliographystyle{aa} 
\bibliography{planets}

\begin{appendix}

\section{Setting initial conditions for MHD simulations}

\label{App:MHDInit}
We describe the method we used to determine the initial conditions for our MHD simulations described in Section \ref{Sec:MHD} 
in order to model disks with a warm corona.
To this aim we start with a corona mildly hotter than the disc's midplane, and determine the vertical density and rotational velocity profiles that correspond to strict hydrostatic and centrifugal equilibria. We then progressively increase the corona's temperature until we reach the desired corona to disc temperature ratio.

We assume that the sound speed is inversely proportional to the spherical radius and has a separable dependence on $r$ and $\theta$:
\begin{equation}
c_s^2(r) = (c_{s0}^0)^2 \left( \frac{r}{r_0} \right)^{-1} g(\theta),
\end{equation}
where   $g(\theta)$  is an even function with respect to $\pi/2$ and $g(\pi/2) = 1$ and $r_0$ is an arbitrary reference radius where the sound speed, at the midplane, is $c_{s0}^0$.
We have considered for $g(\theta)$ the function $T(\tilde \theta)$
defined in Eq.\ref{Tcd}.
We consider a constant disk aspect ratio:
\begin{equation}
h(r) = \frac{c_{s0}(r)}{v_K(r)} = \text{const},
\end{equation}
where $v_K(r) = \sqrt{\frac{GM_*}{r}}$ is the circular Keplerian velocity at midplane, which implies that $c_{s0}^0 \equiv hv_k(r_0)$.

The equations for rotational and vertical equilibrium in spherical coordinates are:

\begin{align}
-\frac{\partial_r (\rho_0 c_s^2)}{\rho_0} + \frac{v_\phi^2}{r} - \frac{GM_*}{r^2} &= 0, \label{eq:A3}\\
-\frac{1}{r} \frac{\partial_\theta (\rho_0 c_s^2)}{\rho_0} + \frac{v_\phi^2}{r} \cot \theta &= 0.\label{eq:A4}
\end{align}

With $\rho_0$ and $c_{s0}(r) = c_{s0}^0 \times \left(\frac{r_0}{r}\right)^{1/2}$ respectively the density and the sound speed at the midplane at radius $r$. We introduce notations similar to those of \citet{2016ApJ...817...19M} and define:
\[
L = \log \rho_0, \quad v = \log \left(\frac{r}{r_0}\right), \quad u = -\log(\sin \theta)
\]

Equations \eqref{eq:A3} and \eqref{eq:A4} become:

\begin{align}
-c_s^2 \partial_v L + c_s^2 + v_\phi^2 - \frac{GM_*}{r} &= 0, \\
c_{s0}^2(r) \frac{\partial_u[\rho_0 g(\theta)]}{\rho_0} + v_\phi^2 &= 0.
\end{align}

Introducing $m \equiv \frac{v_\phi^2}{c_s^2}$, we get:

\begin{align}
-\partial_v L + 1 + m - \frac{GM_*}{rc_s^2} &= 0, \\
c_{s0}^2 g(u) \partial_u L + c_{s0}^2 \partial_u g(u) + v_\phi^2 &= 0.
\end{align}

As per our assumption of a constant aspect ratio, $K \equiv \frac{GM_*}{rc_{s0}^2}$ is a constant and the equations above can be recast as:

\begin{align}
-\partial_v L + 1 + m - \frac{K}{g(u)} &= 0, \\
\partial_u L + \frac{\partial_u g(u)}{g(u)} + m &= 0.
\end{align}

Define $G(u) = \frac{1}{g(u)}$, and $L' = L + \log g(u)$, giving:

\begin{align}
-\partial_v L' + 1 + m - K G(u) &= 0, \\
\partial_u L' + m &= 0.
\end{align}

Let $m' = m - K G(u)$, so:

\begin{align}
-\partial_v L' + 1 + m' &= 0, \label{eq:A13}\\
\partial_u L' + m' + K G(u) &= 0.\label{eq:A14}
\end{align}

The values of $m'$ and $L'$ are determined using a method of characteristics.
We introduce the following combinations of the coordinates $u$ and $v$:

\begin{align}
s &= u + v, \\
s' &= u - v,
\end{align}
so
\begin{align}
\label{Eq:sprime}
u &= \frac{1}{2}(s + s'), \\
v &= \frac{1}{2}(s - s').
\end{align}

Using Eq. \ref{eq:A13} and \ref{eq:A14} we derive:

\[
\partial_s m' = \frac{1}{2}(\partial_u m' + \partial_v m') = 0
\]

Thus, $m'$ depends only on $s'$, 
and its dependence on $s’$ can be determined at the midplane. Considering  a constant $h$ and a midplane density $\rho_{mid}=\frac {\Sigma_0 (r/r_0)^{-\alpha_{\Sigma}-1}}{\sqrt {2\pi}h}$ we obtain from Eq.~\ref{eq:A13}:

\[
\xi + 1 + m' = 0 \Rightarrow m' = -\xi - 1
\]
with $\xi = \alpha_{\Sigma}+1$
Hence:

\begin{equation}
m = h^{-2} G(u) - 1 - \xi \label{eq:A19}
\end{equation}

From Eq.~\eqref{eq:A14}, we get:

\begin{equation}
\partial_u L' + h^{-2} G(u) - 1 - \xi = 0
\end{equation}

Let $W(u) = \int_0^u G(u') \, du'$, then:

\[
L'(u) = L'(0) - h^{-2} W(u) + (1 + \xi)u
\]
hence
\[
L(u) = -\log g(u) + L(0) - h^{-2} W(u) + (1 + \xi)u
\]

Thus the density profile is:

\begin{align}
\rho(u) &= \frac{1}{g(u)} \rho_{\text{mid}} \exp[-h^{-2}W(u) + (1 + \xi)u] \\
&= \rho_{00} \left( \frac{r}{r_0} \right)^{-\xi} (\sin \theta)^{-1 - \xi} G(u) \exp[-h^{-2} W(u)]
\end{align}
with $\rho_{00} = \frac {\Sigma_0}{\sqrt {2\pi}h}$.
Finally, from the the  expression for $m$ (Eq. \ref{eq:A19}),  the azimuthal velocity is:

\begin{align}
v_\phi^2 &= (c_{s0})^2 h^{-2} - (1 + \xi)c_s^2 \\
&= (c_{s0}^0)^2 \left( \frac{r_0}{r} \right) \frac{GM_*}{r_0 (c_{s0}^0)^2} - (1 + \xi)c_s^2 \\
&= \frac{GM_*}{r} - (1 + \xi)c_s^2
\end{align}

\section{MHD simulations diagnostics }

\label{App:MHDDiag}
In the following we provide some diagnostics on our MHD simulations based on the study of  mass transport through the disk in terms or
mass flux or mass accretion (or mass loss through the wind) rates.

The radial mass flux is generated by torques that determine a loss of angular momentum. We follow \cite{Bethune2017} and compute quantities defined in their  Eq.8 and Eq.9 and reported here:

\begin{itemize}
    \item The azimuthal average of a quantity $X$
    is denoted by :$$<X>_{\varphi}= {1\over \Delta \varphi }\int_{0}^{\Delta \varphi} X(r,\theta,\phi,t)d\phi$$

    \item Radial profiles are obtained by vertical integration in the disk domain:
    $$<X>_{\varphi,\theta}= {1\over {2H(r)}}\int _{-H(r)}^{H(r)}<X>_{\varphi}(r,\theta,t)d\theta$$
    \item A density weighted quantity is denoted by:
    $$<X>_{\rho}={<\rho X>_{\varphi,\theta}\over <\rho>_{\varphi,\theta}}$$
    \item The fluctuating Reynold stress tensor is defined by: $$\mathcal{R}\equiv \rho \tilde {\bf v}  \otimes \tilde {\bf v}$$
    where $\tilde {\bf v} = {\bf v}-<v>_{\rho}$.
    \item The Maxwell stress tensor is:
    $$\mathcal { M} \equiv -{\bf B}\otimes {\bf B}$$
    and we call $\mathcal{T} \equiv \mathcal{R}+\mathcal { M}$
    \item We use the same notation as \cite{ShakSun1973} to define $\alpha$ values by normalizing with vertically integrated pressure:
    \begin{equation}
    \label{alpha}
        \alpha^{\mathcal{R}}(r,\theta,t)={<\mathcal{R}>_{\varphi} \over <P>_{\varphi,\theta}} 
    \end{equation}
   
    and similarly for values associated to the magnetic field, denoted $\alpha^{\mathcal{M}}$.
    \end{itemize}

    \begin{figure}
\centering
\includegraphics[width=\hsize]
   {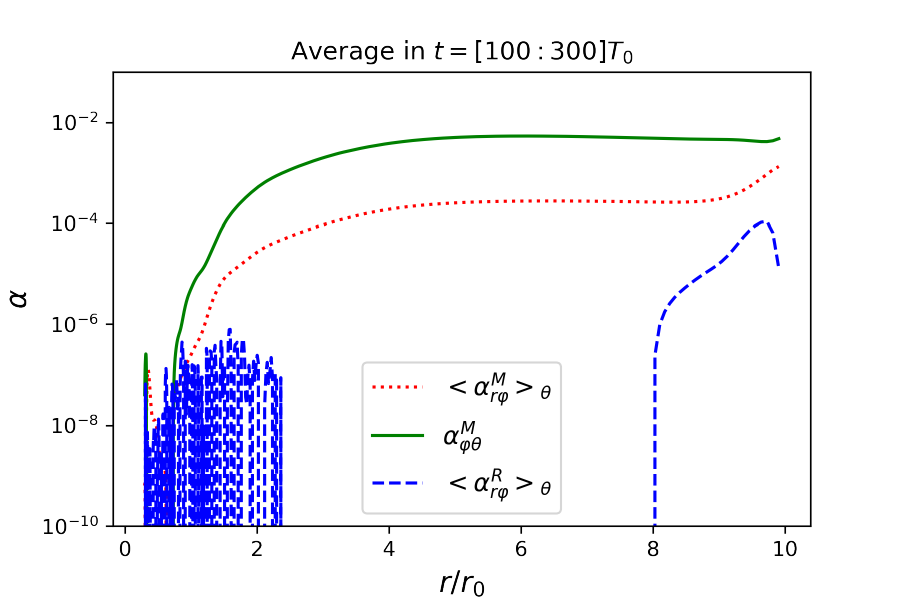}
      \caption{Computation of the $\alpha$ values from Eq.\ref{alpha} averaged on time for the fiducial MHD simulation with $\beta=5\,10^{3}$ and $\Lambda_A=1$.} 
         \label{Fig:AlphaAverage} 
\end{figure}
    
 The radial mass flux 
    can be deduced from the conservation of angular momentum, provided $\Sigma(r) = 2H(r)<\rho>_{\varphi,\theta} $:
    \begin{equation}
        \label{RadialTransport}
        \left\lbrace \begin{array}{lll}
    \Sigma(r) <v_r>_{\rho} & \equiv &  \tau_r + \tau_{z} \\
    \tau_r & \simeq & -{1 \over {r(\partial _r <rv_{\varphi}>_{\varphi,\theta})}}  
    \partial _r (2r^2H(r)<\mathcal{T}_{r,\varphi}>_{\varphi,\theta}) \\
    \tau_{z} & \simeq & - {r \over {[ {\partial _r} <rv_{\varphi}>_{\varphi,\theta}]}}  
    (<\mathcal{T}_{\varphi,\theta}>_{\varphi})_{-H}^{+H} 
    
    \end{array} \right.
    \end{equation}
    where $\tau_r$ gives the mass flux due to the radial component of the angular momentum and $\tau_{z}$ gives the mass transport due to angular momentum extracted at disk surfaces $z=\pm H$

\begin{figure}
\centering
\includegraphics[width=\hsize]
   {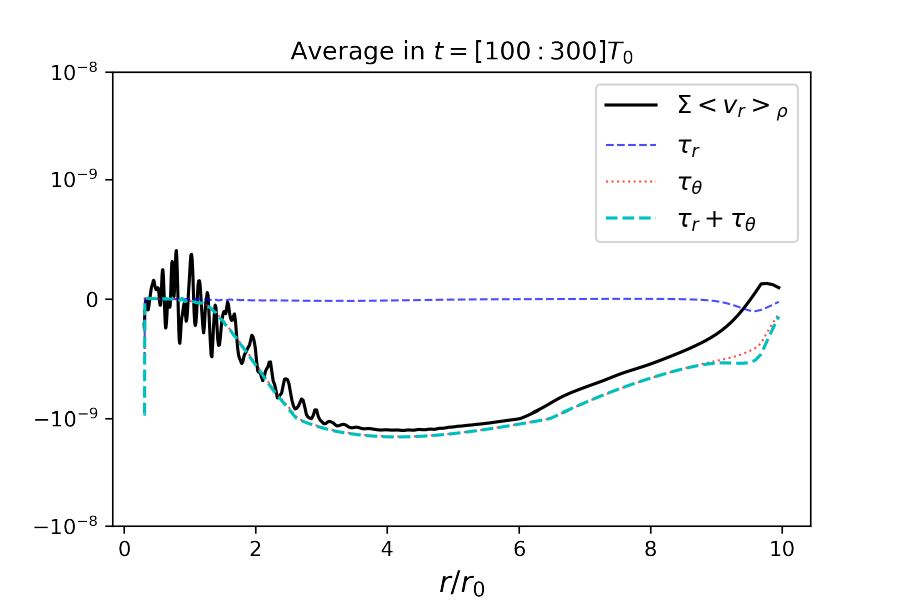}
  \includegraphics[width=\hsize] 
    {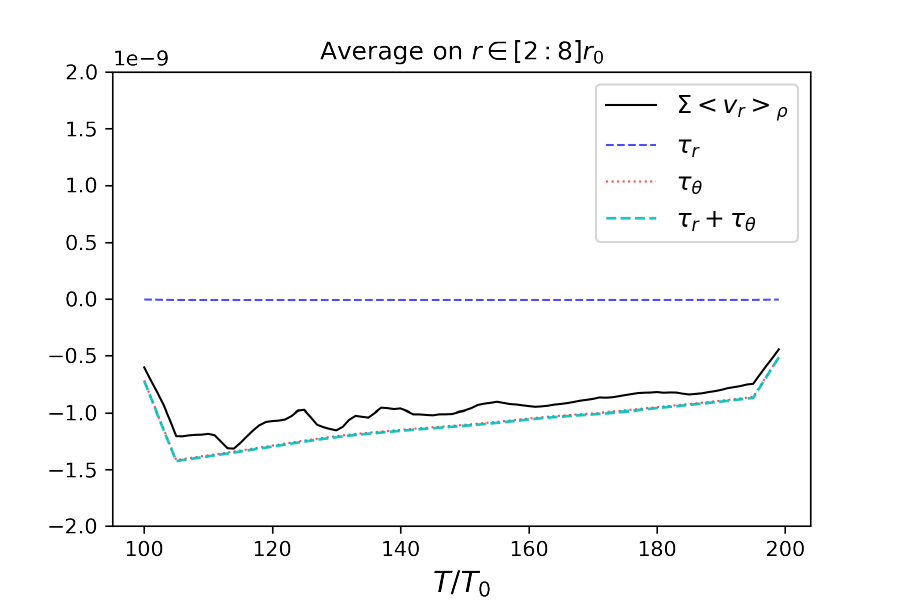}
      \caption{Computation of the radial and vertical contribution to the mass flux (Eq.\ref{RadialTransport}) compared to the measured mass flux (l.h.s. of Eq.\ref{RadialTransport}). Top panel: temporal average, bottom panel: spatial average.}
         \label{Fig:Tau} 
\end{figure}

In figure \ref{Fig:AlphaAverage} we  clearly see that the $\alpha$ values associated to the magnetic field dominate with respect to the Reynold $\alpha$ values.
In fig.\ref{Fig:Tau} we can appreciate that the l.h.s. and the sum of the $\tau$
component of the r.h.s. of the equation follow nicely the same trend both when averaged on time and on space in the interval $r \in [2:8]r_0$.
Moreover, in this radial interval there is  no contribution to the mass flux from the radial transport of the angular momentum. Precisely, let's notice (Fig.\ref{Fig:AlphaAverage}), that $<\alpha_{r\varphi}^{\mathcal{M}}>_{\varphi,\theta}$  is practically constant
for $r \in [2:8]r_0$, and  it is the divergence of this term
(i.e of $\mathcal{T}_{r\varphi}$ in Eq.\ref{RadialTransport})
that contributes  to the radial mass transport. \par
Therefore, the mass flux is completely due to the angular momentum extracted from the disk in the vertical direction (in this case it is
the value of $\mathcal{T}_{\varphi,\theta}$ in Eq.\ref{RadialTransport} at $z=\pm H$ that contributes to $\tau _{\theta}$).\par
This result is in agreement with the one  described in the fiducial simulation  of \cite{Bethune2017} (their Fig.7).

\section{Boundary conditions for MHD simulations}
\label{App:MHDBound}
 In our MHD simulations we have considered the following boundary conditions:
 \begin{itemize}
     \item {\bf Hydro dynamical quantities:}
    \begin{enumerate}
         \item  {\bf inner radial boundary}:  We extrapolate the density and the energy from the active cells into the ghost cells according to the initial power law profile. We extrapolate the azimuthal velocity component of the active cells into the ghost cells according to the  Keplerian profile. The polar component of the velocity is copied from the  active cell into the ghosts cells.
     We close the inner boundary by setting to zero the radial component of the velocity.
        \item {\bf outer radial boundary}: 
        We extrapolate the density and the energy from the active cells into the ghost cells according to the initial power law profile. The azimuthal component of the velocity is extrapolated according to the Keplerian profile. The polar component of the velocity is copied from the  active cell into the ghosts cells. For the radial velocity component we compare the value of the active cell to a threshold value  $\tilde v_r$ proportional  to a fraction of the sound speed. We set the velocity in the ghost cells equal to $\tilde v_r$ if the velocity of the active cell is smaller than the threshold value otherwise we simply copy the value of the active cell into the ghosts.
\item{\bf polar boundaries} {We extrapolate the density and the energy fields according to the hydrostatic vertical equilibrium  defined in Appendix \ref{App:MHDInit}. We extrapolate the azimuthal velocity according to rotational equilibrium (Appendix \ref{App:MHDInit}). We copy the radial and the vertical components of the velocity from the active cells into the ghost cells.}
            \end{enumerate}
       \item {\bf Magnetic field components and EMF}
            
     \begin{enumerate} 
          \item  {\bf inner and outer radial boundary}: 
          We extrapolate linearly the values of the electromotive forces from the active cells into the ghost cells. We copy the azimuthal and the vertical components of the magnetic field from the active cells into the ghosts. We derive the radial component of the magnetic field  by requiring that the divergence of the magnetic field in the ghost cells is zero.
     \item {\bf  polar boundaries} {We extrapolate linearly the values of the electromotive forces from the active cells into the ghost cells. We copy the azimuthal and the radial components of the magnetic field from the active cells into the ghosts. We derive the polar component of the magnetic field  by requiring that the divergence of the magnetic field in the ghost cells is zero.}
     \end{enumerate}
     \end{itemize}
     Finally, we impose a density floor by requiring that the minimum value of the density is given by:
     \begin{equation}
         \label{dens_min}
         \rho_{min} = \frac  {B_{r}^{2}+B_{\theta}^2+B_{\varphi}^2}{4v_{kin}\mu _0}
     \end{equation}
     where $v_{kin}$ is the Keplerian velocity at the inner radial boundary. In other words, we force the Alfv\'en speed to remain
	smaller than few times  the Keplerian velocity at the inner boundary.

\end{appendix}
\end{document}